\pdfoutput=1
\RequirePackage{ifpdf}
\ifpdf 
\documentclass[pdftex]{sigma}
\else
\documentclass{sigma}
\fi

\numberwithin{equation}{section}

\usepackage{tikz}
\usetikzlibrary{positioning}
\usetikzlibrary{patterns}

\newcommand{\wt}{\widetilde}
\newcommand{\R}{\mathbb{R}}
\newcommand{\bigO}{\mathcal{O}}

\begin{document}
\allowdisplaybreaks

\newcommand{\arXivNumber}{2104.00593}

\renewcommand{\thefootnote}{}

\renewcommand{\PaperNumber}{087}

\FirstPageHeading

\ShortArticleName{Semiclassical Trans-Series from the Perturbative Hopf-Algebraic DSEs}

\ArticleName{Semiclassical Trans-Series from the Perturbative\\ Hopf-Algebraic Dyson--Schwinger Equations:\\ $\boldsymbol{\phi^3}$ QFT in 6 Dimensions\footnote{This paper is a~contribution to the Special Issue on Algebraic Structures in Perturbative Quantum Field Theory in honor of Dirk Kreimer for his 60th birthday. The~full collection is available at \href{https://www.emis.de/journals/SIGMA/Kreimer.html}{https://www.emis.de/journals/SIGMA/Kreimer.html}}}

\Author{Michael BORINSKY~$^{\rm a}$, Gerald V. DUNNE~$^{\rm b}$ and Max MEYNIG~$^{\rm b}$}
\AuthorNameForHeading{M.~Borinsky, G.V.~Dunne and M.~Meynig}

\Address{$^{\rm a)}$~Nikhef Theory Group, Amsterdam 1098 XG, The Netherlands}
\EmailD{\href{mailto:michael.borinsky@nikhef.nl}{michael.borinsky@nikhef.nl}}

\Address{$^{\rm b)}$~Department of Physics, University of Connecticut, Storrs CT 06269-3046, USA}
\EmailD{\href{mailto:gerald.dunne@uconn.edu}{gerald.dunne@uconn.edu}, \href{mailto:max.meynig@uconn.edu}{max.meynig@uconn.edu}}

\ArticleDates{Received April 07, 2021, in final form September 16, 2021; Published online September 23, 2021}

\Abstract{We analyze the asymptotically free massless scalar $\phi^3$ quantum field theo\-ry in 6~dimensions, using resurgent asymptotic analysis to find the trans-series solutions which yield the non-perturbative completion of the divergent perturbative solutions to the Kreimer--Connes Hopf-algebraic Dyson--Schwinger equations for the anomalous dimension. This scalar conformal field theory is asymptotically free and has a real Lipatov instanton. In~the Hopf-algebraic approach we find a trans-series having an intricate Borel singularity structure, with three distinct but resonant non-perturbative terms, each repeated in an infi\-nite series. These expansions are in terms of the renormalized coupling. The resonant structure leads to powers of logarithmic terms at higher levels of the trans-series, analogous to logarithmic terms arising from interactions between instantons and anti-instantons, but arising from a~purely perturbative formalism rather than from a semi-classical ana\-lysis.\looseness=1}

\vspace{1mm}

\vspace{1mm}
\Classification{81T15; 81Q15; 34E10}

\renewcommand{\thefootnote}{\arabic{footnote}}
\setcounter{footnote}{0}

\vspace{2mm}

\section{Introduction}
\vspace{1mm}
The seminal work of Kreimer and Connes showed that there is an underlying Hopf-algebraic structure to the renormalization of quantum field theory (QFT)~\cite{Connes:1999zw,Connes:2000fe, Kreimer:1997dp}. This new perspective has led to deep insights into QFT, and also to novel computational methods that have enabled significant progress in higher order perturbative computations~\cite{Bogner:2017xhp,borinsky2018graphs,Borinsky:2020rqs, BoSc21graphical, Broadhurst:1999ys,Broadhurst:2000dq,Clavier:2019sph,Kompaniets:2017yct, Kreimer:2006ua,Kreimer:2006gm,Kruger:2019tas,Panzer:2014caa,Schnetz:2013hqa,Schnetz:2017bko,Schnetz:2016fhy, vanBaalen:2008tc,vanBaalen:2009hu, Yeats:2008zy,yeats2017combinatorial}.
The Hopf-algebraic formulation of QFT is inherently perturbative in nature, so an important open question is to understand how the non-perturbative features of QFT arise naturally within the perturbative Hopf algebra structure.
In a recent paper~\cite{Borinsky:2020vae} we showed how this works for 4 dimensional massless Yukawa theory, using \'Ecalle's theory of resurgent trans-series and alien calculus~\cite{Aniceto:2018bis, costin2008asymptotics,Dorigoni:2014hea,ecalle1981fonctions,mitschi2016divergent,sauzin}. Here we extend this analysis to a conformal field theory: massless scalar $\phi^3$ theory in six dimensional space-time. This QFT has been studied extensively from numerous directions, and has many interesting features, both perturbative and non-perturbative. The theory is asymptotically free for real coupling $g$~\cite{Cardy:1975fz,Cornwall:1995dr,Ma:1975vn,Macfarlane:1974vp}, and has a Yang--Lee edge singularity when $g$ is imaginary~\cite{Fisher:1978pf}. The perturbative beta function and anomalous dimensions have been computed to 4 loop \mbox{order}~\cite{Gracey:2015tta} (and very recently to 5~loop order~\cite{Borinsky:2021jdb,BoSc21graphical}). The perturbative Hopf alge\-bra structure of Dyson--Schwinger equations of this model was formulated in the pioneering papers~\cite{Broadhurst:1999ys,Broadhurst:2000dq}. On the non-perturbative side, this QFT has a real Lipatov instanton when $g$ is real, for which the conventional one-instanton semi-classical analysis~\cite{Brezin:1976vw,lipatov,McKane:2018ocs,ZinnJustin:2002ru} of the fluctuation determinant has been studied~\cite{mckane-thesis,Mckane:1978me}.
Further extensions to multi-dimensional cubic interactions have many interesting applications and implications for conformal quantum field theories in~general~\cite{Benedetti:2020iku,Borinsky:2021jdb, deAlcantaraBonfim:1980pe,deAlcantaraBonfim:1981sy, Fei:2014yja,Giombi:2019upv,Gracey:2015tta,Gracey:2020baa, Gracey:2020tkk, Grinstein:2014xba,Houghton:1978dt}. For other analyses of resurgence properties of~renormalization group and Dyson--Schwinger equations see~\cite{Bellon:2010sf,Bellon:2016mje,Bellon:2020uzi,Bellon:2020qlx, Bellon:2008zz,Bersini:2019axn}.\looseness=1

\looseness=1
Our technical analysis is based on the fundamental result~\cite{Broadhurst:1999ys,Broadhurst:2000dq,Kreimer:2006ua,Kreimer:2006gm} that the Dyson--Schwinger equations have a recursive Hopf-algebraic structure which, when combined with the renormalization group equations describing the anomalous scaling under re-scaling of parameters {and in the absence of vertex renormalization, reduces the problem to a non-linear ordinary differential equation (ODE), where the variable is the {\it renormalized} coupling.
 This Hopf-algebraic approximation goes well beyond the familiar rainbow~\cite{Delbourgo:1996nw} and chain~\cite{Broadhurst:1999ys,Broadhurst:2000dq} approximations to the Dyson--Schwinger equations.
These results cast}
the Hopf algebra renormalization problem in a form in which very high orders of perturbation theory become accessible, and as we show here it also enables direct access to the associated non-perturbative structure. We employ the trans-series approach to the resurgence properties of non-linear differential equations, along the lines of~\cite{costin-imrn,costin1998,costin2008asymptotics}. Our main new result is that the perturbative Hopf algebra formulation encodes a non-perturbative trans-series that involves {powers of} all three {\it trans-monomial elements}: $x$,~${\rm e}^{-1/x}$, and $\log(x)$, all expressed in terms of the renormalized coupling. Moreover, this trans-series has the form of an all-orders multi-instanton expansion, and the logarithms appear with the characteristic structure of logarithmic terms arising from the interaction of instantons and anti-instantons.\footnote{This logarithmic structure does not occur for the 4 dimensional Yukawa model studied in~\cite{Borinsky:2020vae}.} Logarithmic terms are familiar in semi-classical computations~\cite{alvarez_howls_silverstone,ddp,ddp+, Dunne:2012ae,Dunne:2013ada,Dunne:2016nmc,Lapedes:1981tz,Marino:2012zq,Misumi:2015dua, ZinnJustin:1979db,ZinnJustin:2004ib,ZinnJustin:2004ib+}, and have been studied in differential equations where resonant Borel singularities $\pm A$ interact~\cite{Aniceto:2011nu,damburg,Garoufalidis:2010ya}, but here we find a quite different resonant Borel structure, with three resonant singularities of the same sign yet in integer multiples.
All this non-perturbative information is encoded in the original perturbative Hopf-algebraic formulation, which at first sight makes no explicit mention of instantons, let alone interactions between instantons and anti-instantons.

\section[Perturbative Hopf-algebraic analysis of massless phi3 theory in 6 dimensions]
{Perturbative Hopf-algebraic analysis\\ of massless $\boldsymbol{\phi^3}$ theory in 6 dimensions}

In this paper we analyze the massless scalar $\phi^3$ theory in 6 dimensional spacetime. This is the critical dimension in which the theory is asymptotically free~\cite{Macfarlane:1974vp} and in which it has a Lipatov instanton~\cite{lipatov,mckane-thesis,Mckane:1978me}. We analyze the non-perturbative features arising in the Hopf-algebraic approach of~\cite{Broadhurst:1999ys,Broadhurst:2000dq,Kreimer:2006ua,Kreimer:2006gm}. The Lagrangian density is
\begin{gather*}
{\mathcal L}= \frac{1}{2} (\partial_\mu \phi )^2-\frac{g}{3!} \phi^3.
\end{gather*}

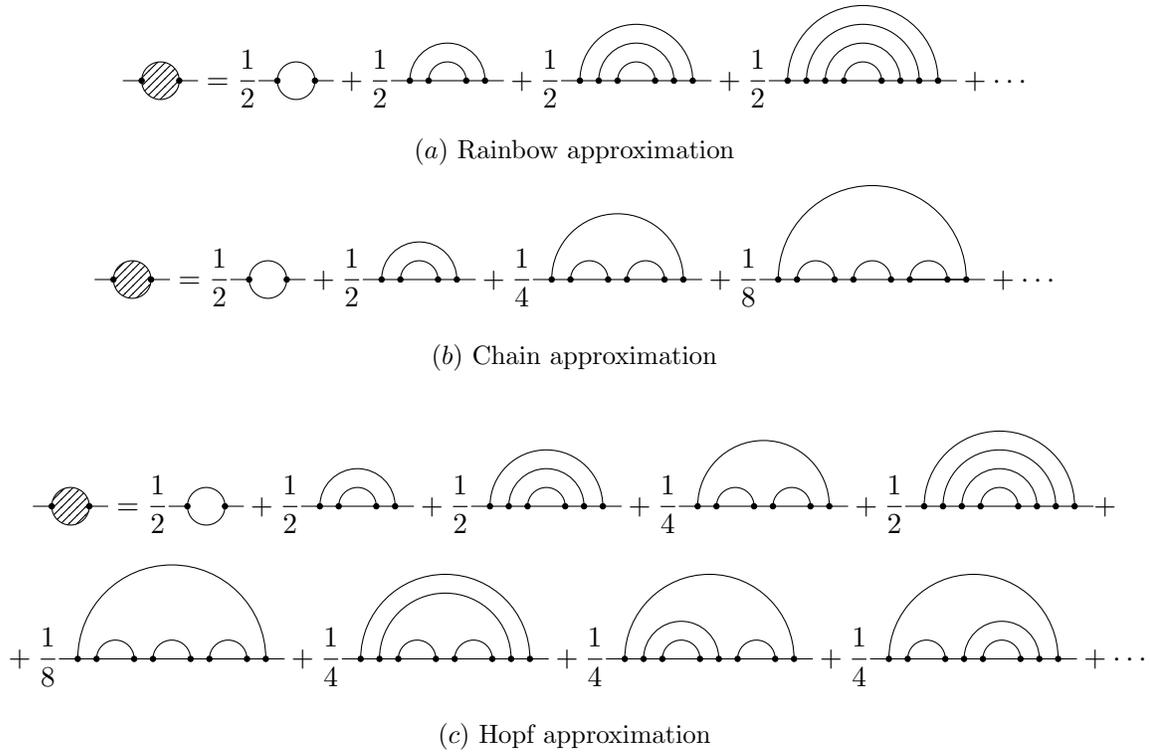
\begin{figure}
$$
\begin{tikzpicture}[baseline={([yshift=-.6ex]current bounding box.center)}] \coordinate (in); \coordinate[right=.25 of in] (v1); \coordinate[right=.25 of v1] (m); \coordinate[right=.25 of m] (v2); \coordinate[right=.25 of v2] (out); \filldraw[preaction={fill,white},pattern=north east lines] (m) circle(.25); \filldraw (v1) circle(1pt); \filldraw (v2) circle(1pt); \draw (in) -- (v1); \draw (v2) -- (out); \end{tikzpicture}=
\frac12
\begin{tikzpicture}[baseline={([yshift=-.6ex]current bounding box.center)}] \coordinate (in); \coordinate[right=.25 of in] (v1); \coordinate[right=.25 of v1] (vm); \coordinate[right=.25 of vm] (v2); \coordinate[right=.25 of v2] (out); \filldraw (v1) circle(1pt); \filldraw (v2) circle(1pt); \draw (in) -- (v1); \draw (v2) -- (out); \draw (vm) circle (.25); \end{tikzpicture}
+
\frac12
\begin{tikzpicture}[baseline={([yshift=-.6ex]current bounding box.center)}] \coordinate (in); \coordinate[right=.25 of in] (v1); \coordinate[right=.25 of v1] (v2); \coordinate[right=.25 of v2] (m); \coordinate[right=.25 of m] (v3); \coordinate[right=.25 of v3] (v4); \coordinate[right=.25 of v4] (out); \draw[white] (v1) arc (-180:0:.5); \filldraw (v1) circle(1pt); \filldraw (v2) circle(1pt); \filldraw (v3) circle(1pt); \filldraw (v4) circle(1pt); \draw (in) -- (v1); \draw (v1) -- (v2); \draw (v2) -- (v3); \draw (v3) -- (v4); \draw (v4) -- (out); \draw (v1) arc (180:0:.5); \draw (v2) arc (180:0:.25); \end{tikzpicture}
+
\frac12
\begin{tikzpicture}[baseline={([yshift=-.6ex]current bounding box.center)}] \coordinate (in); \coordinate[right=.25 of in] (v1); \coordinate[right=.25 of v1] (v2); \coordinate[right=.25 of v2] (v3); \coordinate[right=.25 of v3] (m); \coordinate[right=.25 of m] (v4); \coordinate[right=.25 of v4] (v5); \coordinate[right=.25 of v5] (v6); \coordinate[right=.25 of v6] (out); \draw[white] (v1) arc (-180:0:.75); \filldraw (v1) circle(1pt); \filldraw (v2) circle(1pt); \filldraw (v3) circle(1pt); \filldraw (v4) circle(1pt); \filldraw (v5) circle(1pt); \filldraw (v6) circle(1pt); \draw (in) -- (v1); \draw (v1) -- (v2); \draw (v2) -- (v3); \draw (v3) -- (v4); \draw (v4) -- (v5); \draw (v5) -- (v6); \draw (v6) -- (out); \draw (v1) arc (180:0:.75); \draw (v2) arc (180:0:.5); \draw (v3) arc (180:0:.25); \end{tikzpicture}
+
\frac12
\begin{tikzpicture}[baseline={([yshift=-.6ex]current bounding box.center)}] \coordinate (in); \coordinate[right=.25 of in] (v1); \coordinate[right=.25 of v1] (v2); \coordinate[right=.25 of v2] (v3); \coordinate[right=.25 of v3] (v4); \coordinate[right=.25 of v4] (m); \coordinate[right=.25 of m] (v5); \coordinate[right=.25 of v5] (v6); \coordinate[right=.25 of v6] (v7); \coordinate[right=.25 of v7] (v8); \coordinate[right=.25 of v8] (out); \draw[white] (v1) arc (-180:0:1); \filldraw (v1) circle(1pt); \filldraw (v2) circle(1pt); \filldraw (v3) circle(1pt); \filldraw (v4) circle(1pt); \filldraw (v5) circle(1pt); \filldraw (v6) circle(1pt); \filldraw (v7) circle(1pt); \filldraw (v8) circle(1pt); \draw (in) -- (v1); \draw (v1) -- (v2); \draw (v2) -- (v3); \draw (v3) -- (v4); \draw (v4) -- (v5); \draw (v5) -- (v6); \draw (v6) -- (v7); \draw (v7) -- (v8); \draw (v8) -- (out); \draw (v1) arc (180:0:1); \draw (v2) arc (180:0:.75); \draw (v3) arc (180:0:.5); \draw (v4) arc (180:0:.25); \end{tikzpicture}
+
\cdots \hfill\vspace{-3ex}$$
$$\text{\small ($a$) Rainbow approximation}$$
$$
\begin{tikzpicture}[baseline={([yshift=-.6ex]current bounding box.center)}] \coordinate (in); \coordinate[right=.25 of in] (v1); \coordinate[right=.25 of v1] (m); \coordinate[right=.25 of m] (v2); \coordinate[right=.25 of v2] (out); \filldraw[preaction={fill,white},pattern=north east lines] (m) circle(.25); \filldraw (v1) circle(1pt); \filldraw (v2) circle(1pt); \draw (in) -- (v1); \draw (v2) -- (out); \end{tikzpicture}=
\frac12
\begin{tikzpicture}[baseline={([yshift=-.6ex]current bounding box.center)}] \coordinate (in); \coordinate[right=.25 of in] (v1); \coordinate[right=.25 of v1] (vm); \coordinate[right=.25 of vm] (v2); \coordinate[right=.25 of v2] (out); \filldraw (v1) circle(1pt); \filldraw (v2) circle(1pt); \draw (in) -- (v1); \draw (v2) -- (out); \draw (vm) circle (.25); \end{tikzpicture}
+
\frac{1}{2}
\begin{tikzpicture}[baseline={([yshift=-.6ex]current bounding box.center)}] \coordinate (in); \coordinate[right=.25 of in] (v1); \coordinate[right=.25 of v1] (v2); \coordinate[right=.25 of v2] (m); \coordinate[right=.25 of m] (v3); \coordinate[right=.25 of v3] (v4); \coordinate[right=.25 of v4] (out); \draw[white] (v1) arc (-180:0:.5); \filldraw (v1) circle(1pt); \filldraw (v2) circle(1pt); \filldraw (v3) circle(1pt); \filldraw (v4) circle(1pt); \draw (in) -- (v1); \draw (v1) -- (v2); \draw (v2) -- (v3); \draw (v3) -- (v4); \draw (v4) -- (out); \draw (v1) arc (180:0:.5); \draw (v2) arc (180:0:.25); \end{tikzpicture}
+
\frac{1}{4}
\begin{tikzpicture}[baseline={([yshift=-.6ex]current bounding box.center)}] \coordinate (in); \coordinate[right=.25 of in] (v1); \coordinate[right=.25 of v1] (v2); \coordinate[right=.25 of v2] (m1); \coordinate[right=.25 of m1] (v3); \coordinate[right=.25 of v3] (v4); \coordinate[right=.25 of v4] (m2); \coordinate[right=.25 of m2] (v5); \coordinate[right=.25 of v5] (v6); \coordinate[right=.25 of v6] (out); \draw[white] (v1) arc (-180:0:.875); \filldraw (v1) circle(1pt); \filldraw (v2) circle(1pt); \filldraw (v3) circle(1pt); \filldraw (v4) circle(1pt); \filldraw (v5) circle(1pt); \filldraw (v6) circle(1pt); \draw (in) -- (v1); \draw (v1) -- (v2); \draw (v2) -- (v3); \draw (v3) -- (v4); \draw (v4) -- (v5); \draw (v5) -- (v6); \draw (v6) -- (out); \draw (v1) arc (180:0:.875); \draw (v2) arc (180:0:.25); \draw (v4) arc (180:0:.25); \end{tikzpicture}
+
\frac{1}{8}
\begin{tikzpicture}[baseline={([yshift=-.6ex]current bounding box.center)}] \coordinate (in); \coordinate[right=.25 of in] (v1); \coordinate[right=.25 of v1] (v2); \coordinate[right=.25 of v2] (m1); \coordinate[right=.25 of m1] (v3); \coordinate[right=.25 of v3] (v4); \coordinate[right=.25 of v4] (m2); \coordinate[right=.25 of m2] (v5); \coordinate[right=.25 of v5] (v6); \coordinate[right=.25 of v6] (m3); \coordinate[right=.25 of m3] (v7); \coordinate[right=.25 of v7] (v8); \coordinate[right=.25 of v8] (out); \draw[white] (v1) arc (-180:0:1.25); \filldraw (v1) circle(1pt); \filldraw (v2) circle(1pt); \filldraw (v3) circle(1pt); \filldraw (v4) circle(1pt); \filldraw (v5) circle(1pt); \filldraw (v6) circle(1pt); \filldraw (v7) circle(1pt); \filldraw (v8) circle(1pt); \draw (in) -- (v1); \draw (v1) -- (v2); \draw (v2) -- (v3); \draw (v3) -- (v4); \draw (v4) -- (v5); \draw (v5) -- (v6); \draw (v6) -- (v7); \draw (v7) -- (v8); \draw (v6) -- (out); \draw (v1) arc (180:0:1.25); \draw (v2) arc (180:0:.25); \draw (v4) arc (180:0:.25); \draw (v6) arc (180:0:.25); \end{tikzpicture}
+
\cdots \hfill\vspace{-4ex}$$
$${\text{\small ($b$) Chain approximation}}\vspace{-2ex}$$
\\
$$
\begin{tikzpicture}[baseline={([yshift=-.6ex]current bounding box.center)}] \coordinate (in); \coordinate[right=.25 of in] (v1); \coordinate[right=.25 of v1] (m); \coordinate[right=.25 of m] (v2); \coordinate[right=.25 of v2] (out); \filldraw[preaction={fill,white},pattern=north east lines] (m) circle(.25); \filldraw (v1) circle(1pt); \filldraw (v2) circle(1pt); \draw (in) -- (v1); \draw (v2) -- (out); \end{tikzpicture}=
\frac12
\begin{tikzpicture}[baseline={([yshift=-.6ex]current bounding box.center)}] \coordinate (in); \coordinate[right=.25 of in] (v1); \coordinate[right=.25 of v1] (vm); \coordinate[right=.25 of vm] (v2); \coordinate[right=.25 of v2] (out); \filldraw (v1) circle(1pt); \filldraw (v2) circle(1pt); \draw (in) -- (v1); \draw (v2) -- (out); \draw (vm) circle (.25); \end{tikzpicture}
+
\frac12
\begin{tikzpicture}[baseline={([yshift=-.6ex]current bounding box.center)}] \coordinate (in); \coordinate[right=.25 of in] (v1); \coordinate[right=.25 of v1] (v2); \coordinate[right=.25 of v2] (m); \coordinate[right=.25 of m] (v3); \coordinate[right=.25 of v3] (v4); \coordinate[right=.25 of v4] (out); \draw[white] (v1) arc (-180:0:.5); \filldraw (v1) circle(1pt); \filldraw (v2) circle(1pt); \filldraw (v3) circle(1pt); \filldraw (v4) circle(1pt); \draw (in) -- (v1); \draw (v1) -- (v2); \draw (v2) -- (v3); \draw (v3) -- (v4); \draw (v4) -- (out); \draw (v1) arc (180:0:.5); \draw (v2) arc (180:0:.25); \end{tikzpicture}
+
\frac12
\begin{tikzpicture}[baseline={([yshift=-.6ex]current bounding box.center)}] \coordinate (in); \coordinate[right=.25 of in] (v1); \coordinate[right=.25 of v1] (v2); \coordinate[right=.25 of v2] (v3); \coordinate[right=.25 of v3] (m); \coordinate[right=.25 of m] (v4); \coordinate[right=.25 of v4] (v5); \coordinate[right=.25 of v5] (v6); \coordinate[right=.25 of v6] (out); \draw[white] (v1) arc (-180:0:.75); \filldraw (v1) circle(1pt); \filldraw (v2) circle(1pt); \filldraw (v3) circle(1pt); \filldraw (v4) circle(1pt); \filldraw (v5) circle(1pt); \filldraw (v6) circle(1pt); \draw (in) -- (v1); \draw (v1) -- (v2); \draw (v2) -- (v3); \draw (v3) -- (v4); \draw (v4) -- (v5); \draw (v5) -- (v6); \draw (v6) -- (out); \draw (v1) arc (180:0:.75); \draw (v2) arc (180:0:.5); \draw (v3) arc (180:0:.25); \end{tikzpicture}
+
\frac14
\begin{tikzpicture}[baseline={([yshift=-.6ex]current bounding box.center)}] \coordinate (in); \coordinate[right=.25 of in] (v1); \coordinate[right=.25 of v1] (v2); \coordinate[right=.25 of v2] (m1); \coordinate[right=.25 of m1] (v3); \coordinate[right=.25 of v3] (v4); \coordinate[right=.25 of v4] (m2); \coordinate[right=.25 of m2] (v5); \coordinate[right=.25 of v5] (v6); \coordinate[right=.25 of v6] (out); \draw[white] (v1) arc (-180:0:.875); \filldraw (v1) circle(1pt); \filldraw (v2) circle(1pt); \filldraw (v3) circle(1pt); \filldraw (v4) circle(1pt); \filldraw (v5) circle(1pt); \filldraw (v6) circle(1pt); \draw (in) -- (v1); \draw (v1) -- (v2); \draw (v2) -- (v3); \draw (v3) -- (v4); \draw (v4) -- (v5); \draw (v5) -- (v6); \draw (v6) -- (out); \draw (v1) arc (180:0:.875); \draw (v2) arc (180:0:.25); \draw (v4) arc (180:0:.25); \end{tikzpicture}
+
\frac12
\begin{tikzpicture}[baseline={([yshift=-.6ex]current bounding box.center)}] \coordinate (in); \coordinate[right=.25 of in] (v1); \coordinate[right=.25 of v1] (v2); \coordinate[right=.25 of v2] (v3); \coordinate[right=.25 of v3] (v4); \coordinate[right=.25 of v4] (m); \coordinate[right=.25 of m] (v5); \coordinate[right=.25 of v5] (v6); \coordinate[right=.25 of v6] (v7); \coordinate[right=.25 of v7] (v8); \coordinate[right=.25 of v8] (out); \draw[white] (v1) arc (-180:0:1); \filldraw (v1) circle(1pt); \filldraw (v2) circle(1pt); \filldraw (v3) circle(1pt); \filldraw (v4) circle(1pt); \filldraw (v5) circle(1pt); \filldraw (v6) circle(1pt); \filldraw (v7) circle(1pt); \filldraw (v8) circle(1pt); \draw (in) -- (v1); \draw (v1) -- (v2); \draw (v2) -- (v3); \draw (v3) -- (v4); \draw (v4) -- (v5); \draw (v5) -- (v6); \draw (v6) -- (v7); \draw (v7) -- (v8); \draw (v8) -- (out); \draw (v1) arc (180:0:1); \draw (v2) arc (180:0:.75); \draw (v3) arc (180:0:.5); \draw (v4) arc (180:0:.25); \end{tikzpicture}
+\vspace{-3ex}$$
$${} +
\frac18
\begin{tikzpicture}[baseline={([yshift=-.6ex]current bounding box.center)}] \coordinate (in); \coordinate[right=.25 of in] (v1); \coordinate[right=.25 of v1] (v2); \coordinate[right=.25 of v2] (m1); \coordinate[right=.25 of m1] (v3); \coordinate[right=.25 of v3] (v4); \coordinate[right=.25 of v4] (m2); \coordinate[right=.25 of m2] (v5); \coordinate[right=.25 of v5] (v6); \coordinate[right=.25 of v6] (m3); \coordinate[right=.25 of m3] (v7); \coordinate[right=.25 of v7] (v8); \coordinate[right=.25 of v8] (out); \draw[white] (v1) arc (-180:0:1.25); \filldraw (v1) circle(1pt); \filldraw (v2) circle(1pt); \filldraw (v3) circle(1pt); \filldraw (v4) circle(1pt); \filldraw (v5) circle(1pt); \filldraw (v6) circle(1pt); \filldraw (v7) circle(1pt); \filldraw (v8) circle(1pt); \draw (in) -- (v1); \draw (v1) -- (v2); \draw (v2) -- (v3); \draw (v3) -- (v4); \draw (v4) -- (v5); \draw (v5) -- (v6); \draw (v6) -- (v7); \draw (v7) -- (v8); \draw (v6) -- (out); \draw (v1) arc (180:0:1.25); \draw (v2) arc (180:0:.25); \draw (v4) arc (180:0:.25); \draw (v6) arc (180:0:.25); \end{tikzpicture}
+
\frac14
\begin{tikzpicture}[baseline={([yshift=-.6ex]current bounding box.center)}] \coordinate (in); \coordinate[right=.25 of in] (v0); \coordinate[right=.25 of v0] (v1); \coordinate[right=.25 of v1] (v2); \coordinate[right=.25 of v2] (m1); \coordinate[right=.25 of m1] (v3); \coordinate[right=.25 of v3] (v4); \coordinate[right=.25 of v4] (m2); \coordinate[right=.25 of m2] (v5); \coordinate[right=.25 of v5] (v6); \coordinate[right=.25 of v6] (v7); \coordinate[right=.25 of v7] (out); \draw[white] (v1) arc (-180:0:1.125); \filldraw (v0) circle(1pt); \filldraw (v1) circle(1pt); \filldraw (v2) circle(1pt); \filldraw (v3) circle(1pt); \filldraw (v4) circle(1pt); \filldraw (v5) circle(1pt); \filldraw (v6) circle(1pt); \filldraw (v7) circle(1pt); \draw (in) -- (v1); \draw (v1) -- (v2); \draw (v2) -- (v3); \draw (v3) -- (v4); \draw (v4) -- (v5); \draw (v5) -- (v6); \draw (v6) -- (out); \draw (v0) arc (180:0:1.125); \draw (v1) arc (180:0:.875); \draw (v2) arc (180:0:.25); \draw (v4) arc (180:0:.25); \end{tikzpicture}%
+
\frac14
\begin{tikzpicture}[baseline={([yshift=-.6ex]current bounding box.center)}] \coordinate (in); \coordinate[right=.25 of in] (v1); \coordinate[right=.25 of v1] (v2); \coordinate[right=.25 of v2] (v3); \coordinate[right=.25 of v3] (m1); \coordinate[right=.25 of m1] (v4); \coordinate[right=.25 of v4] (v5); \coordinate[right=.25 of v5] (v6); \coordinate[right=.25 of v6] (m2); \coordinate[right=.25 of m2] (v7); \coordinate[right=.25 of v7] (v8); \coordinate[right=.25 of v8] (out); \draw[white] (v1) arc (-180:0:1.125); \filldraw (v1) circle(1pt); \filldraw (v2) circle(1pt); \filldraw (v3) circle(1pt); \filldraw (v4) circle(1pt); \filldraw (v5) circle(1pt); \filldraw (v6) circle(1pt); \filldraw (v7) circle(1pt); \filldraw (v8) circle(1pt); \draw (in) -- (v1); \draw (v1) -- (v2); \draw (v2) -- (v3); \draw (v3) -- (v4); \draw (v4) -- (v5); \draw (v5) -- (v6); \draw (v6) -- (out); \draw (v1) arc (180:0:1.125); \draw (v2) arc (180:0:.5); \draw (v3) arc (180:0:.25); \draw (v6) arc (180:0:.25); \end{tikzpicture}%
+
\frac14
\begin{tikzpicture}[baseline={([yshift=-.6ex]current bounding box.center)}] \coordinate (in); \coordinate[left=.25 of in] (v1); \coordinate[left=.25 of v1] (v2); \coordinate[left=.25 of v2] (v3); \coordinate[left=.25 of v3] (m1); \coordinate[left=.25 of m1] (v4); \coordinate[left=.25 of v4] (v5); \coordinate[left=.25 of v5] (v6); \coordinate[left=.25 of v6] (m2); \coordinate[left=.25 of m2] (v7); \coordinate[left=.25 of v7] (v8); \coordinate[left=.25 of v8] (out); \draw[white] (v1) arc (0:-180:1.125); \filldraw (v1) circle(1pt); \filldraw (v2) circle(1pt); \filldraw (v3) circle(1pt); \filldraw (v4) circle(1pt); \filldraw (v5) circle(1pt); \filldraw (v6) circle(1pt); \filldraw (v7) circle(1pt); \filldraw (v8) circle(1pt); \draw (in) -- (v1); \draw (v1) -- (v2); \draw (v2) -- (v3); \draw (v3) -- (v4); \draw (v4) -- (v5); \draw (v5) -- (v6); \draw (v6) -- (out); \draw (v1) arc (0:180:1.125); \draw (v2) arc (0:180:.5); \draw (v3) arc (0:180:.25); \draw (v6) arc (0:180:.25); \end{tikzpicture}%
+
\cdots\vspace{-4ex}$$
$${\text{\small ($c$) Hopf approximation}}$$
\caption{A comparison of the diagrams included in various approximations for the self-energy of~the mass\-less $\phi^3$ theory: ($a$) the rainbow approximation; ($b$) the chain approximation; and ($c$) the Hopf approximation studied in this paper. The relevant symmetry factors are also indicated. Note that the three approximations agree up to two-loop order, but differ at higher orders. At the third loop order, the Hopf approximation is the sum of the other two. Beyond third order, the Hopf approximation includes new classes of diagrams which are not present in either the rainbow or the chain approximation.}
 \label{fig:rainbow_chain_approximation}
\end{figure}

{\samepage\noindent
As in~\cite{Broadhurst:1999ys,Broadhurst:2000dq} we consider the renormalized scalar self-energy
\begin{align*}
\Pi\big(q^2\big) :=
\begin{tikzpicture}[baseline={([yshift=-.6ex]current bounding box.center)}] \coordinate (in); \coordinate[right=.25 of in] (v1); \coordinate[right=.25 of v1] (m); \coordinate[right=.25 of m] (v2); \coordinate[right=.25 of v2] (out); \filldraw[preaction={fill,white},pattern=north east lines] (m) circle(.25); \filldraw (v1) circle(1pt); \filldraw (v2) circle(1pt); \draw (in) -- (v1); \draw (v2) -- (out); \end{tikzpicture}
\end{align*}}\noindent
and take all propagator self-insertions into account. This {Hopf-algebraic} approach is depicted by the Dyson--Schwinger equation
\begin{align}
\begin{aligned}
\label{eq:pic_dse}
\begin{tikzpicture}[baseline={([yshift=-.6ex]current bounding box.center)}] \coordinate (in); \coordinate[right=.25 of in] (v1); \coordinate[right=.25 of v1] (m); \coordinate[right=.25 of m] (v2); \coordinate[right=.25 of v2] (out); \filldraw[preaction={fill,white},pattern=north east lines] (m) circle(.25); \filldraw (v1) circle(1pt); \filldraw (v2) circle(1pt); \draw (in) -- (v1); \draw (v2) -- (out); \end{tikzpicture}
&=
\frac12
\begin{tikzpicture}[baseline={([yshift=-.6ex]current bounding box.center)}] \coordinate (in); \coordinate[right=.25 of in] (v1); \coordinate[right=.25 of v1] (vm); \coordinate[right=.25 of vm] (v2); \coordinate[right=.25 of v2] (out); \filldraw (v1) circle(1pt); \filldraw (v2) circle(1pt); \draw (in) -- (v1); \draw (v2) -- (out); \draw (vm) circle (.25); \end{tikzpicture}
+
\begin{tikzpicture}[baseline={([yshift=-.6ex]current bounding box.center)}] \coordinate (in); \coordinate[right=.25 of in] (v1); \coordinate[right=.25 of v1] (v2); \coordinate[right=.25 of v2] (m); \coordinate[right=.25 of m] (v3); \coordinate[right=.25 of v3] (v4); \coordinate[right=.25 of v4] (out); \draw[white] (v1) arc (-180:0:.5); \filldraw[preaction={fill,white},pattern=north east lines] (m) circle(.25); \filldraw (v1) circle(1pt); \filldraw (v2) circle(1pt); \filldraw (v3) circle(1pt); \filldraw (v4) circle(1pt); \draw (in) -- (v1); \draw (v1) -- (v2); \draw (v3) -- (v4); \draw (v4) -- (out); \draw (v1) arc (180:0:.5); \end{tikzpicture}%
+
\begin{tikzpicture}[baseline={([yshift=-.6ex]current bounding box.center)}] \coordinate (in); \coordinate[right=.25 of in] (v1); \coordinate[right=.25 of v1] (v2); \coordinate[right=.25 of v2] (m1); \coordinate[right=.25 of m1] (v3); \coordinate[right=.25 of v3] (v4); \coordinate[right=.25 of v4] (m2); \coordinate[right=.25 of m2] (v5); \coordinate[right=.25 of v5] (v6); \coordinate[right=.25 of v6] (out); \draw[white] (v1) arc (-180:0:.5); \draw[white] (v4) arc (-180:0:.5); \filldraw[preaction={fill,white},pattern=north east lines] (m1) circle(.25); \filldraw[preaction={fill,white},pattern=north east lines] (m2) circle(.25); \draw[white] (v1) arc (-180:0:.875); \filldraw (v1) circle(1pt); \filldraw (v2) circle(1pt); \filldraw (v3) circle(1pt); \filldraw (v4) circle(1pt); \filldraw (v5) circle(1pt); \filldraw (v6) circle(1pt); \draw (in) -- (v1); \draw (v1) -- (v2); \draw (v3) -- (v4); \draw (v5) -- (v6); \draw (v6) -- (out); \draw (v1) arc (180:0:.875); \end{tikzpicture}%
+
\cdots
-
\text{subtractions}
\end{aligned}
\end{align}
with the appropriate BPHZ subtractions indicated. Another way to describe the relevant set of graphs is to start with the one-loop graph $\begin{tikzpicture}[baseline={([yshift=-.6ex]current bounding box.center)}] \coordinate (in); \coordinate[right=.25 of in] (v1); \coordinate[right=.25 of v1] (vm); \coordinate[right=.25 of vm] (v2); \coordinate[right=.25 of v2] (out); \filldraw (v1) circle(1pt); \filldraw (v2) circle(1pt); \draw (in) -- (v1); \draw (v2) -- (out); \draw (vm) circle (.25); \end{tikzpicture}$ and add all possible iterated and multiple insertions of this graph into one of the propagators.
Figure~\ref{fig:rainbow_chain_approximation} shows the resulting low order dia\-grams and compares this Hopf expansion with two other common approximations to the Dyson--Schwinger equations: the rainbow approximation~\cite{Delbourgo:1996nw} and the chain approximation~\cite{Broadhurst:1999ys,Broadhurst:2000dq}.
The Hopf expansion includes a much larger class of diagrams than either the rainbow or the chain approximation, and leads to a much richer non-perturbative structure. The differences between these approximations is discussed further below, in Sections \ref{sec:perturbative} and \ref{sec:borel}.\footnote{{The effect of including also the vertex corrections will be addressed in future work.}}

The pictorial Dyson--Schwinger equation \eqref{eq:pic_dse} corresponds to the integral equation
\begin{gather*}
\Pi\big(q^2\big) = \frac{a}{\pi^3}
\int {\rm d}^6 k\,\frac{1}{ (q+k)^2} \bigg( \frac{1}{k^2}
+ \frac{1}{k^2} \Pi\big(k^2\big) \frac{1}{k^2}
+ \frac{1}{k^2} \Pi\big(k^2\big) \frac{1}{k^2} \Pi\big(k^2\big) \frac{1}{k^2}
+ \cdots \bigg)
\\[.5ex] \hphantom{\Pi\big(q^2\big) =}
{}- \text{subtractions},
\end{gather*}
where the natural coupling expansion parameter is
\begin{gather*}
a:=\frac{g^2}{(4\pi)^3}.
\end{gather*}
It is convenient to extract a factor of $q^2$ by defining the function $\wt{\Pi}\big(q^2\big)$, such that
$\Pi\big(q^2\big) = q^2 \wt{\Pi}\big(q^2\big)$,
and the integral equation reduces to
\begin{gather*}
q^2 \wt{\Pi}\big(q^2\big) = \frac{a}{\pi^3} \int {\rm d}^6 k\, \frac{1}{k^2 (q+k)^2 \big(1 - \wt{\Pi}\big(k^2\big)\big)}
-\text{subtractions}.
\end{gather*}
The BPHZ subtractions are chosen such that the momentum subtraction renormalization condition $\wt{\Pi}\big(\mu^2\big) =0$ is satisfied.

The anomalous dimension is defined in the momentum subtraction scheme as
\begin{gather*}
\gamma(a):=\frac{\rm d}{{\rm d}\ln q^2}\ln \big(1-\wt{\Pi}\big(q^2\big)\big)\bigg|_{q^2=\mu^2} .
\end{gather*}
Broadhurst and Kreimer~\cite{Broadhurst:1999ys,Broadhurst:2000dq} showed that the Hopf-algebraic anomalous dimension satisfies the following nonlinear ordinary differential equation:
\begin{gather}
 0 = 8 a^3 \gamma(a) \big(\gamma(a)^2 \gamma''' (a)+\gamma '(a)^3+4 \gamma(a) \gamma '(a) \gamma''(a)\big)\nonumber
\\ \hphantom{ 0 =}
 {}+4 a^2 \gamma(a) \big(2 (\gamma(a)-3) \gamma (a) \gamma ''(a)+(\gamma (a)-6) \gamma '(a)^2\big)\nonumber
\\ \hphantom{ 0 =}
{} +2 a \gamma (a) \big(2 \gamma (a)^2+6 \gamma (a)+11\big) \gamma '(a)-\gamma (a) (\gamma (a)+1) (\gamma (a)+2) (\gamma (a)+3)-a.\label{eq:ode}
\end{gather}
This equation is {\it quartic} in the anomalous dimension $\gamma(a)$, and {\it third order} in derivatives with respect to the renormalized coupling $a$. Contrast this with the massless Yukawa theory in 4 dimensions where the corresponding nonlinear equation for the anomalous dimension
is {\it quadratic} in $\gamma(a)$, and {\it first order} in derivatives with respect to $a$~\cite{Borinsky:2020vae,Broadhurst:1999ys,Broadhurst:2000dq}. We therefore expect the $\phi^3$ theory to have a richer perturbative and non-perturbative structure, as we demonstrate explicitly in this paper.

In this massless theory, the full self-energy can be expanded formally in powers of $L\equiv \ln \frac{q^2}{\mu^2}$, or in powers of the renormalized coupling $a$~\cite{Kreimer:2006ua,yeats2017combinatorial}:
\begin{gather*}
\wt{\Pi}\big(q^2\big)=-\sum_{j=1}^\infty \gamma_j(a) L^j=-\sum_{j=1}^\infty c_j(L) a^j,\\
L\equiv \ln \frac{q^2}{\mu^2}.
\end{gather*}
The first term, $\gamma_1(a)$, is just the anomalous dimension $\gamma(a)$, and all subsequent higher coefficients are expressed recursively in term of $\gamma_1(a)$~\cite{Kreimer:2006ua,yeats2017combinatorial}:
\begin{gather*}
\gamma_k(a)=\frac{1}{k} \gamma_{1}(a)(1-2 a \partial_a) \gamma_{k-1}(a), \qquad k\geq 2.
\end{gather*}
Therefore the basic trans-series structure of $\gamma(a)$ is inherited by all the subsequent $\gamma_k(a)$, and hence also by the self-energy $\tilde{\Pi}\big(q^2\big)$.
The unique perturbative solution in $\R[[a]]$ to the nonlinear ODE~(\ref{eq:ode}) can be generated straightforwardly with a formal perturbative ansatz~\cite{Broadhurst:1999ys,Broadhurst:2000dq}
\begin{gather}
 \gamma (a) := \sum_{n=1}^\infty (-1)^n \frac{A_n}{6^{2n-1}} a^n
 =-\frac{a}{6}+11\frac{a^2}{6^3}-376\frac{a^3}{6^5}+20241 \frac{a^4}{6^7} - 1427156 \frac{a^5}{6^9}\nonumber
 \\ \hphantom{\gamma (a) := \sum_{n=1}^\infty (-1)^n \frac{A_n}{6^{2n-1}} a^n =}
 {}+ 121639250\frac{a^6}{6^{11}}-12007003824\frac{a^7}{6^{13}} +\cdots.
 \label{eq:an}
\end{gather}
The coefficients are all rational, and the normalization of the coefficients was chosen in~\cite{Broadhurst:1999ys,Broadhurst:2000dq} to make the $A_n$ integer-valued. The first few integers $A_n$ are
\begin{gather*}
 \big\{1,11,376,20241,1427156,121639250,12007003824,1337583507153,165328009728652,
 \\ \qquad
 22404009743110566,3299256277254713760, \dots \big\}.
\end{gather*}
This sequence is listed in the OEIS~\cite{oeis} as entry \href{https://oeis.org/A051862}{A051862}.
There is currently no known combinatorial interpretation of the integer $A_n$ coefficients, in contrast to the 4d massless Yukawa model analyzed in~\cite{Borinsky:2020vae, Broadhurst:1999ys,Broadhurst:2000dq}, where the corresponding perturbative expansion is the generating function for connected chord diagrams~\cite{Courtiel:2019dnq,Mahmoud:2020vww,Marie:2012cc, yeats2017combinatorial}.

\vspace{1mm}
\section[Asymptotics of the perturbative solution of the Hopf-algebraic Dyson--Schwinger equation]
{Asymptotics of the perturbative solution \\of the Hopf-algebraic Dyson--Schwinger equation}
\label{sec:perturbative}

Given the ODE (\ref{eq:ode}), it is straightforward to generate recursively very high orders of the perturbative expansion (\ref{eq:an}).\footnote{We thank David Broadhurst for also providing a list of the first 2000 integer coefficients $A_n$.}
Via a simple ratio test combined with high-order Richardson extra\-polation~\cite{bender-book} we can experimentally deduce that the leading large order growth is given by the classical gamma function, multiplied by a power of $12$:
\begin{gather}\label{eq:pert-growth1}
A_n = \frac{S_1}{6} 12^n \Gamma\bigg(n+\frac{23}{12}\bigg) \big(1 + \bigO\left(n^{-1}\right) \big), \qquad \text{for}\quad n\to \infty.
\end{gather}
Therefore the perturbative expansion in (\ref{eq:an}) is a factorially divergent series.
The overall coefficient $S_1$, the Stokes constant, can be determined to very high precision:
\begin{gather}
S_1 \approx 0.08759555290917912448379544742126299062738801740682153692058109
\dots.
\label{eq:S1}
\end{gather}
The normalization factor $\frac16$ was chosen for later convenience. $S_1$ does not appear to be a simple recognizable number.

\begin{figure}[htb]
\centering
\includegraphics[scale=.7]{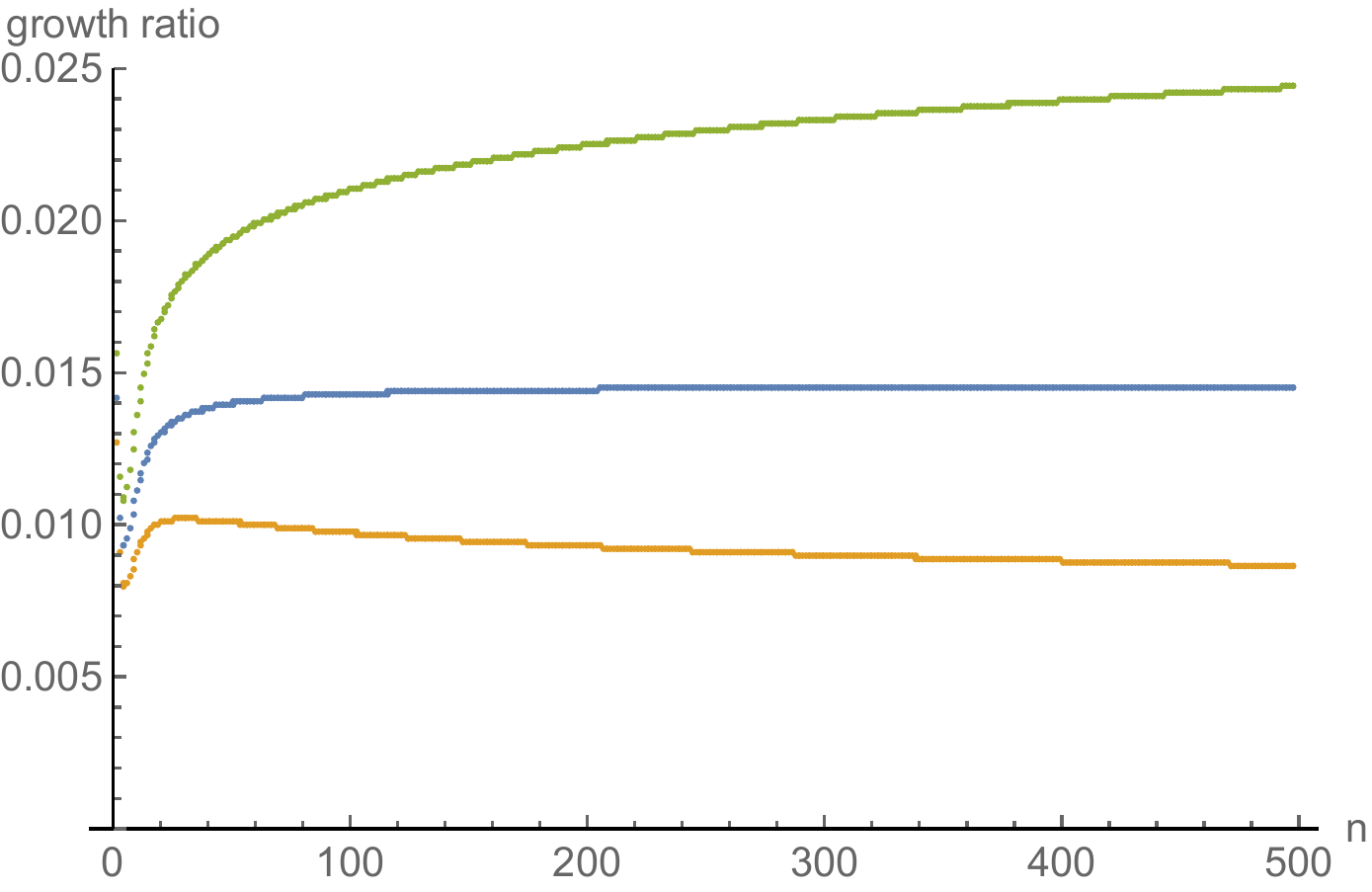}
\caption{The ratio of the perturbative $A_n$ coefficients in (\ref{eq:an}) to the leading order growth $12^n \Gamma\left(n+\frac{23}{12}\right)$ (blue), compared with other nearby offsets: $12^n \Gamma\left(n+\frac{24}{12}\right)$ (gold), and $12^n \Gamma\left(n+\frac{22}{12}\right)$ (green).
The origin of the exact factorial offset $\frac{23}{12}$ is explained analytically in Section \ref{sec:trans}.}
\label{fig:leading-an}
\end{figure}

The leading large $n$ factorial dependence in (\ref{eq:pert-growth1}) is close to, but importantly different from, the apparent $A_n\sim 12^n \Gamma(n+2)$ large order growth estimated in~\cite{Broadhurst:1999ys,Broadhurst:2000dq} based on the first 30 terms. The rational offset $23/12$ of the argument of the Gamma function in (\ref{eq:pert-growth1}) follows analytically from a trans-series analysis (see Section \ref{sec:trans}) of the Dyson--Schwinger equation (\ref{eq:ode}).\footnote{{See also~\cite{Bellon:2010sf}.}}
Figure~\ref{fig:leading-an} shows a numerical illustration of this offset parameter.

It is interesting to compare the Hopf-algebraic perturbative expansion (\ref{eq:an}) with two simpler perturbative approximations to the Dyson--Schwinger equations: the rainbow and the chain approximations, as depicted in Figure~\ref{fig:rainbow_chain_approximation}.
The rainbow approximation~\cite{Broadhurst:1999ys,Broadhurst:2000dq,Delbourgo:1996nw} yields a simple closed-form algebraic expression for the anomalous dimension, which has a {\it convergent} expansion (for ease of comparison, we use the same normalization convention as in (\ref{eq:an})):
\begin{gather*}
\begin{split}
&\gamma_{\rm rainbow}(a):= \sum_{n=1}^\infty (-1)^n \frac{A^{\rm rainbow}_n}{6^{2n-1}} a^n
\\ &\hphantom{\gamma_{\rm rainbow}(a)}
{}=-\frac{a}{6}+11 \frac{a^2}{6^3}-206 \frac{a^3}{6^5}+ 4711 \frac{a^4}{6^7}- 119762\frac{a^5}{6^9} +3251262 \frac{a^6}{6^{11}}+\cdots
\\ &\hphantom{\gamma_{\rm rainbow}(a)}
{}= \frac{1}{2}\Big(3-\sqrt{5+4\sqrt{1+a}}\Big).
\end{split}
\end{gather*}
On the other hand, the chain approximation~\cite{Broadhurst:1999ys,Broadhurst:2000dq} yields a {\it divergent} perturbative expansion of the anomalous dimension (once again with the same normalization convention as in (\ref{eq:an})):
\begin{gather}
\gamma_{\rm chain}(a)
:= \sum_{n=1}^\infty (-1)^n \frac{A^{\rm chain}_n}{6^{2n-1}} a^n
\label{eq:chain0}
\\ \hphantom{\gamma_{\rm chain}(a)}
{}=-\frac{a}{6}+11 \frac{a^2}{6^3}-170 \frac{a^3}{6^5}+ 3450 \frac{a^4}{6^7}-87864 \frac{a^5}{6^9}+ 2715720 \frac{a^6}{6^{11}}+ \cdots
\label{eq:chaina}
\\ \hphantom{\gamma_{\rm chain}(a)}
{}= 3 \sum_{n=1}^\infty (-1)^n \Gamma(n)\bigg(\frac{1}{6^n}-\frac{2}{12^n}+\frac{1}{18^n}\bigg)a^n.
\label{eq:chain}
\end{gather}
Note that the first two terms of the rainbow, chain and Hopf approximations all agree, since they involve the same diagrams, but that the $a^3$ coefficient differs. This is because at this order the relevant diagrams differ, as illustrated in Figure~\ref{fig:rainbow_chain_approximation}. Furthermore, note that the Hopf-algebraic~$a^3$ coefficient in (\ref{eq:an}) is $-376/6^5=-206/6^5-170/6^5$, equal to the sum of the rainbow and chain contributions at this order, as can be understood diagrammatically from Figure~\ref{fig:rainbow_chain_approximation}. This also
illustrates the fact that the Hopf-algebraic analysis incorporates a larger class of diagrams than either the rainbow or chain approximations at this order and beyond. It is also interesting to note that for the first 6 terms it appears that the rainbow approximation coefficients are growing more rapidly than those of the chain approximation, but eventually the factorially divergent growth of the chain approximation coefficients overtakes the growth of the coefficients of the convergent rainbow approximation.

\begin{figure}[htb]
\centering
\includegraphics[scale=.7]{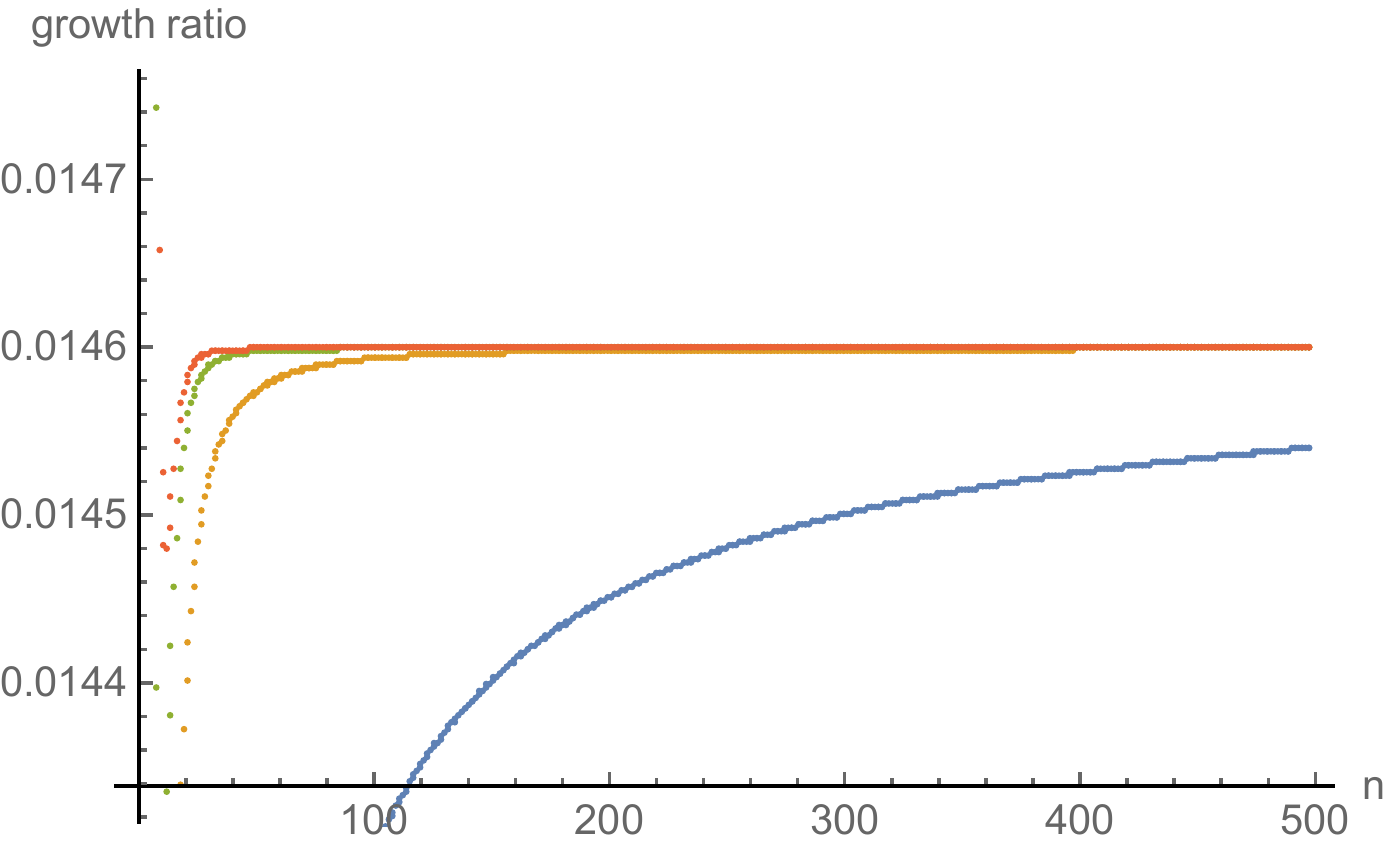}
\caption{The ratio of the perturbative Hopf $A_n$ coefficients in (\ref{eq:an}) to the leading order growth, $12^n \Gamma\left(n+\frac{23}{12}\right)$ (blue). This blue curve is a zoomed-in view of the blue curve in Figure~\ref{fig:leading-an}. The other curves in this figure include successively the first 3 subleading power-law corrections in (\ref{eq:pert-growth2}) (gold, green, red). The asymptotic value is equal to the constant $S_1/6$ in (\ref{eq:S1}).}
\label{fig:subleading-an}
\end{figure}

We use 500 Hopf expansion coefficients in (\ref{eq:an}) to experimentally extract subleading power-law corrections to the leading factorial growth in (\ref{eq:pert-growth1}). The methodology is explained in Appen\-dix~\ref{sec:app}:
\begin{gather}
A_n\sim \frac{S_1}{6} 12^{n} \Gamma\bigg(n+\frac{23}{12}\bigg)\Bigg({1}\!-\frac{{\frac{97}{48}}}{ \left(n+\frac{11}{12}\right) }\!-\frac{{\frac{53917}{13824} } }{ \left(n-\frac{1}{12}\right) \left(n+\frac{11}{12}\right)}
\!-\frac{{\frac{3026443}{221184}} }{\left(n-\frac{13}{12}\right) \left(n-\frac{1}{12}\right) \left(n+\frac{11}{12}\right)} \nonumber
\\ \hphantom{A_n\sim}
 {}
 -\frac{{\frac{32035763261}{382205952}} }{\left(n-\frac{25}{12}\right)\left(n-\frac{13}{12}\right) \left(n-\frac{1}{12}\right) \left(n+\frac{11}{12}\right)}
 - \cdots\Bigg)
 +\cdots , \qquad n\to\infty.
\label{eq:pert-growth2}
\end{gather}
See Figure~\ref{fig:subleading-an} for numerical illustrations of the precision of these subleading power-law corrections to the leading factorial large-order growth.
The first few subleading coefficients in (\ref{eq:pert-growth2}) are
\begin{gather}\label{eq:pert-growth3}
 \bigg\{ 1,-\frac{97}{48},-\frac{53917}{13824},-\frac{3026443}{221184},-\frac{32035763261}{382205952} ,\dots \bigg\}.
\end{gather}
The physical significance of these subleading coefficients in (\ref{eq:pert-growth2})--(\ref{eq:pert-growth3}) is discussed below~-- see Section \ref{sec:seed0}.

The final ellipsis in (\ref{eq:pert-growth2}) refers to further {\it exponentially} suppressed subleading corrections, beyond the {\it power-law} subleading corrections indicated inside the parentheses. These exponentially small corrections cannot be resolved by simple ratio tests and Richardson extrapolations. However, these exponentially small corrections are resolved by the Borel analysis in Section~\ref{sec:borel}, and by the trans-series analysis in Section \ref{sec:trans}. Physically, they correspond to ``higher instanton'' non-perturbative terms, while the expansion in~(\ref{eq:pert-growth2}) corresponds to just the leading ``one-instanton'' term: the leading factor characterizes the one-instanton, while the subleading power-law corrections in (\ref{eq:pert-growth2}) encode the perturbative fluctuations about the single instanton.
Note that while the chain approximation expansion (\ref{eq:chain0})--(\ref{eq:chain}) is also factorially divergent, there is a simple closed-form expression for the expansion coefficients.
Using the same normalization of the coefficients as in the Hopf expansion (\ref{eq:an}), we see that
the chain coefficients defined in (\ref{eq:chain0})--(\ref{eq:chain}) also grow factorially fast:
\begin{eqnarray}
A_n^{\rm chain} =\frac{1}{2}\cdot 6^n \Gamma(n)\bigg(1-\frac{2}{2^n}+\frac{1}{3^n}\bigg).
\label{eq:chain-an}
\end{eqnarray}
This makes it clear that the growth of the chain coefficients is {\it slower} than that of the Hopf coefficients in (\ref{eq:pert-growth2}), because of the factor $6^n$ instead of $12^n$. The argument of the leading factorial factor, $\Gamma(n)$, is also different. The exact expression (\ref{eq:chain-an}) also shows that there are {\it exponentially smaller} corrections to the {\it leading} factorial growth, $A_n^{\rm chain}\sim -\frac{1}{2} 6^n \Gamma(n)$, encoded in the $1/2^n$ and $1/3^n$ correction terms in (\ref{eq:chain-an}). The significance of these exponentially smaller corrections will become clear in the Borel analysis of the next section.

\section[Borel analysis of the perturbative expansion of the anomalous dimension]
{Borel analysis of the perturbative expansion \\of the anomalous dimension}
\label{sec:borel}

In this section we use Borel methods, combined with conformal maps, to analyze in more detail the structure of the formal perturbative solution (\ref{eq:an}) to the
Hopf-algebraic Dyson--Schwinger equation (\ref{eq:ode}). Remarkably, the complicated-looking nonlinear equation~(\ref{eq:ode}) can be factored:
\begin{gather}
 \bigg[{G}(x)\bigg(2 x \frac{\rm d}{{\rm d}x}-1\bigg)-1\bigg]
 \bigg[{G}(x) \bigg(2 x \frac{\rm d}{{\rm d}x}-1\bigg)-2\bigg]
 \bigg[{G}(x) \bigg(2 x \frac{\rm d}{{\rm d}x}-1\bigg)-3\bigg]{G}(x)\nonumber
 \\ \qquad
 {} =-3x.
 \label{eq:ode2}
 \end{gather}
Here we have defined
 \begin{gather}
 G(x):=\gamma(-3 x)
 \label{eq:g}
 \end{gather}
 rescaling the variable as $x:=- \frac{a}{3}$, to account for the alternating sign and the power-law growth factor, $12^n/6^{2n}=1/3^n$, coming from (\ref{eq:an}) and the leading growth in (\ref{eq:pert-growth1}).\footnote{{In terms of the physical expansion variable $a=-3x=\frac{g^2}{(4\pi)^3}$, the change of sign from $a$ to $x$ is of interest in the context of the Lee--Yang edge singularity for the theory where $g$ is pure imaginary~\cite{deAlcantaraBonfim:1981sy, Fisher:1978pf,mckane-thesis,Mckane:1978me}. It is also of interest for the inclusion of vertex diagrams, as will be discussed in future work. For these reasons we choose to work here in terms of the variable~$x$, where exponential terms are most easily identified.}}
Then the formal perturbative series in (\ref{eq:an}) becomes
\begin{gather}
G^{\text{pert}}(x):= 6 \sum_{n=1}^\infty \frac{A_n}{12^n} x^n
\label{eq:g-exp}
\\ \hphantom{G^{\text{pert}}(x)}
{}\sim
\frac{x}{2}+\frac{11 x^2}{24}+\frac{47 x^3}{36}+\frac{2249 x^4}{384}+ \frac{356789 x^5}{10368} +\frac{60819625 x^6}{248832}
+\cdots.
 \label{eq:g-exp2}
\end{gather}
With this scaling, the coefficients of $x^n$ in $G^{\text{pert}}(x)$ have leading growth that is purely factorial, $\sim \Gamma\left(n+\frac{23}{12}\right)$, with the exponential factor $12^n$ in (\ref{eq:pert-growth1}) scaled out.

We define the corresponding Borel transform:
\begin{gather}
 \mathcal B^{\rm pert} (t):=
 6\sum_{n=1}^\infty \frac{A_n}{12^n} \frac{t^n}{n!}.
 \label{eq:BorelPert}
\end{gather}
The formal perturbative series for $G(x)$ is recovered by the Laplace transform:\footnote{{This formal integral expression is of course to be understood in the Borel--\'Ecalle sense~\cite{costin2008asymptotics,Dorigoni:2014hea,ecalle1981fonctions,mitschi2016divergent,sauzin}.}}
\begin{gather*}
G^{\rm pert} (x)=\frac{1}{x} \int_0^\infty {\rm d}t \,{\rm e}^{-t/x} \mathcal B^{\rm pert} (t).
\label{eq:laplace}
\end{gather*}%
\begin{figure}[htb!]
\centering
\includegraphics[width=100mm]{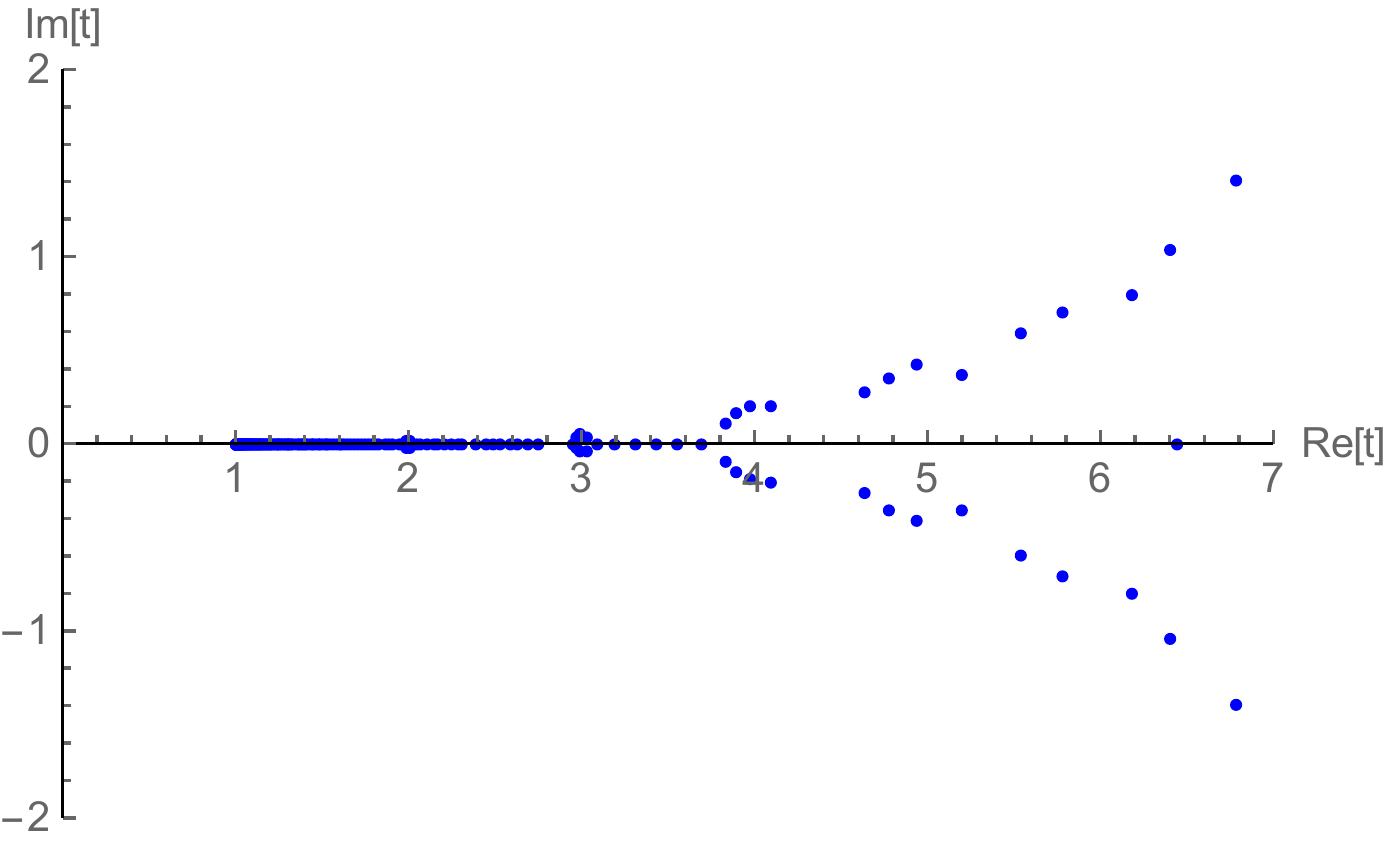}
 \caption{The blue points show the poles of the diagonal Pad\'e approximant of the Borel transform (\ref{eq:BorelPert}) of the Hopf-algebraic perturbative series \eqref{eq:g-exp}.
 There is a clear accumulation of Pad\'e poles to the point $t=1^+$, indicating the presence of a branch point there. There are also hints of further singularities at integer multiples of the leading singularity, with diminishing resolution due to the finite order of Pad\'e. This structure is resolved more clearly in Figures \ref{fig:pcb-z-poles} and \ref{fig:PertPCBLogPlot} after using a conformal map.}
 \label{fig:PertBorelPoles}
\end{figure}

With this choice of scaling,
the Borel transform (\ref{eq:BorelPert}) has radius of convergence equal to~$1$. To probe the leading Borel singularity more closely, a natural first step is to use Pad\'e approxi\-mants~\cite{bender-book}. Figure~\ref{fig:PertBorelPoles} shows the poles of the diagonal Pad\'e approximant to $\mathcal B^{\rm pert} (t)$, truncated after 500 terms. Recall that a Pad\'e approximant, being a rational approximation of the function as a ratio of polynomials, has only pole singularities. But a Pad\'e approximant of a function with branch points represents a branch point as an accumulation point of poles~\cite{Costin:2020hwg,Costin:2020pcj,stahl}. Figure~\ref{fig:PertBorelPoles} suggests that the leading Borel singularity is at $t=1$, consistent with the radius of convergence being $1$. Furthermore, the accumulation of Pad\'e poles to $t=1^+$ suggests that this leading singularity is a branch point rather than a pole. This reveals a drawback of the Pad\'e approximant: in attempting to represent a branch cut by a line of poles along the interval $t\in [1, \infty)$, accumulating to $t=1$, it obscures the possible existence of other branch points along this same line. Since the Borel singularities correspond physically to non-perturbative ``instanton'' terms, with the leading singularity at $t=1$ being the ``one-instanton'' singularity, in a nonlinear problem such as this we expect this leading singularity to be repeated at integer multiples, corresponding to the ``multi-instanton'' terms. A~closer look at Figure~\ref{fig:PertBorelPoles} hints at the possible existence of other singularities at integer values along the positive $t$ axis, but is unable to resolve them clearly.

Fortunately there is a simple way to cure this problem. We can resolve these higher Borel singularities using a conformal mapping method that has been widely used (albeit for different reasons) in the physics literature~\cite{caliceti,caprini,jentschura2001improved,ZinnJustin:2002ru}.
The idea is to make a conformal map\footnote{Significantly higher precision can be obtained with a uniformization map~\cite{Costin:2020pcj}, which is particularly useful if fewer perturbative terms are available.} from the cut Borel $t$ plane, based on the leading singularity at $t=+1$, to the unit disk in the $z$ plane:
\begin{gather}
z=\frac{1-\sqrt{1-t}}{1+\sqrt{1-t}} , \qquad t=\frac{4z}{(1+z)^2}.
\label{eq:conformal1}
\end{gather}
Note that this conformal map does not require knowledge of the {\it nature} of the leading Borel singularity, just its location. In fact, the high precision of the conformal mapping step means that it can be used to iteratively refine the location of the leading singularity if it is only known approximately~\cite{Costin:2020hwg,Costin:2020pcj}.
By construction, any further singularities on the positive Borel $t$ axis, beyond the leading one at $t=+1$, are mapped to the unit circle in the $z$ plane. The conformal map takes $t=+1$ to $z=+1$, and $t=+2$ to $z=\pm {\rm i}$, representing both sides of the first cut, and so on.

This {\it Pad\'e--conformal--Borel} procedure is as follows:
\begin{enumerate}\itemsep=0pt
\item Re-expand the Borel transform $\mathcal B\big(4z/(1+z)^2\big)$ about $z=0$ inside the unit disk of the conformally mapped $z$ plane, to the same order as the original expansion in the Borel $t$ plane.
\item Make a diagonal Pad\'e approximant to the resulting truncated series in~$z$.
\item Find the singularities (poles) in $z$ of this Pad\'e approximant. Higher branch points in the~$t$~plane will now be separated as points of accumulation of poles to the unit circle in the~$z$~plane.
\item To obtain an accurate analytic continuation in the original Borel $t$ plane, especially near the branch points, map this Pad\'e approximant in $z$ back to the Borel $t$ plane with the inverse conformal map in (\ref{eq:conformal1}): the resulting analytic continuation of the Borel transform is denoted $\mathcal{PCB}(t)$.
\end{enumerate}

\begin{figure}[htb!]
\centering
\includegraphics[width=90mm]{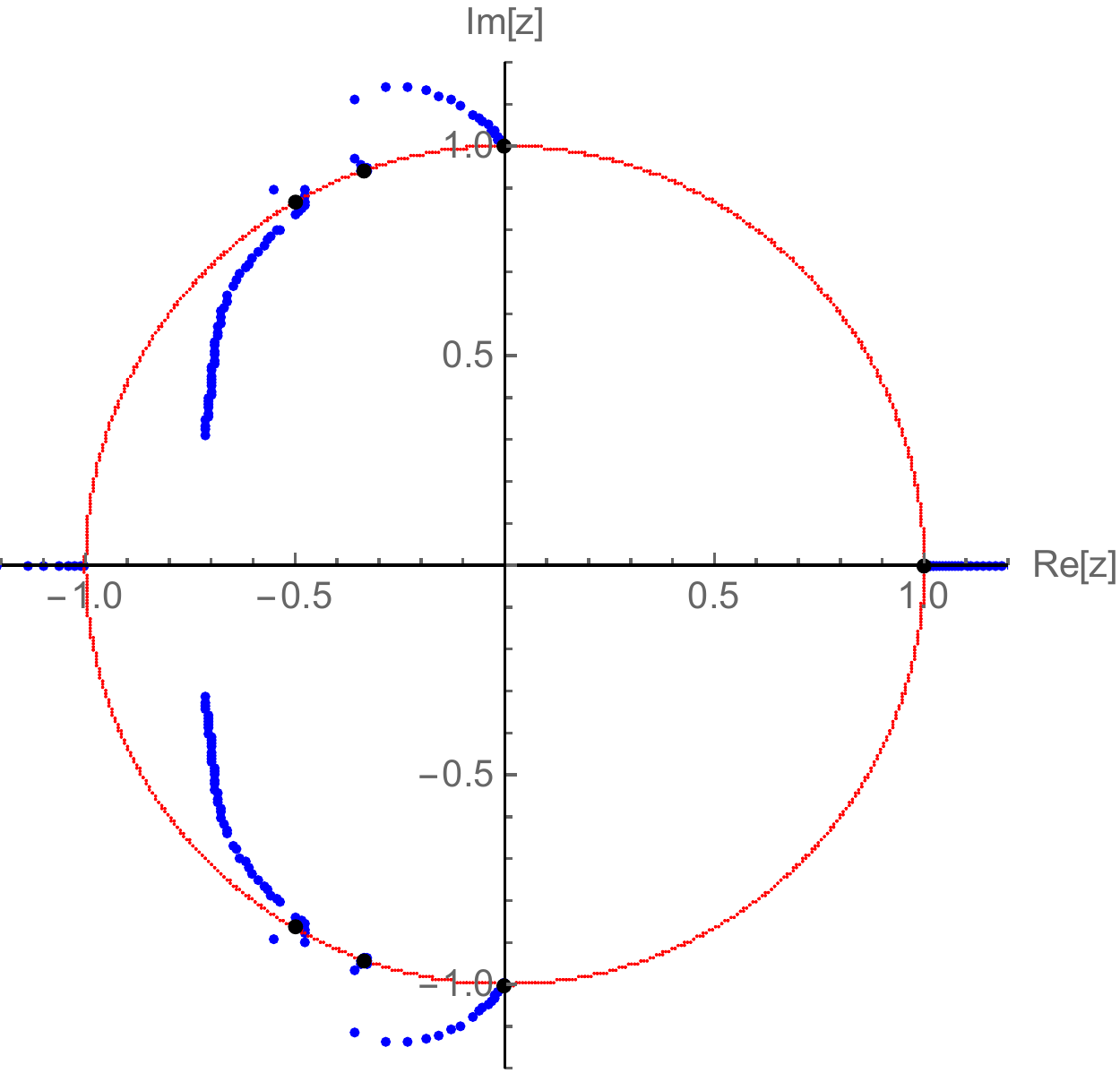}
 \caption{The blue points indicate poles in the complex $z$ plane of the diagonal Pad\'e approximant of the Borel transform (\ref{eq:BorelPert}) after the conformal mapping (\ref{eq:conformal1}), followed by re-expansion about $z=0$. The~poles accumulate to the conformal map images (black points) of the points $t=+1, +2, +3, +4$ in the Borel $t$ plane. This illustrates how the conformal map separates and resolves the genuine physical Borel singularities.}
 \label{fig:pcb-z-poles}
\end{figure}

Figure~\ref{fig:pcb-z-poles} plots the $z$ poles of the Pad\'e approximant of the conformally mapped series (in step~3, above), demonstrating that multiple branch points, and their associated branch cuts, have been separated and resolved by the conformal map to the conformal $z$ plane~\cite{Costin:2020hwg,Costin:2020pcj}. All~these branch points lie on the real positive Borel axis, in the interval $t\in [1, \infty)$.
The singularity at $z=+1$ is the conformal map image of the leading singularity at $t=+1$. The singularities at $z=\pm {\rm i}$ correspond to the conformal map image of the two-instanton Borel singularity at $t=+2$, on either side of the branch cut. We can further resolve a third singularity at $z=-\frac{1}{3}\pm \frac{2\sqrt{2}}{3} {\rm i}$, which is the conformal map image of the three-instanton Borel singularity at $t=+3$, once again on either side of the branch cut. There is also weaker evidence of a further singularity at the conformal image of $t=+4$, corresponding to $z={\rm e}^{\pm 2\pi {\rm i}/3}$. This can be resolved by taking more terms in the original perturbative expansion, therefore allowing a higher order Pad\'e approximation.

\begin{figure}[h!]
	\centering
	\includegraphics[width=90mm]{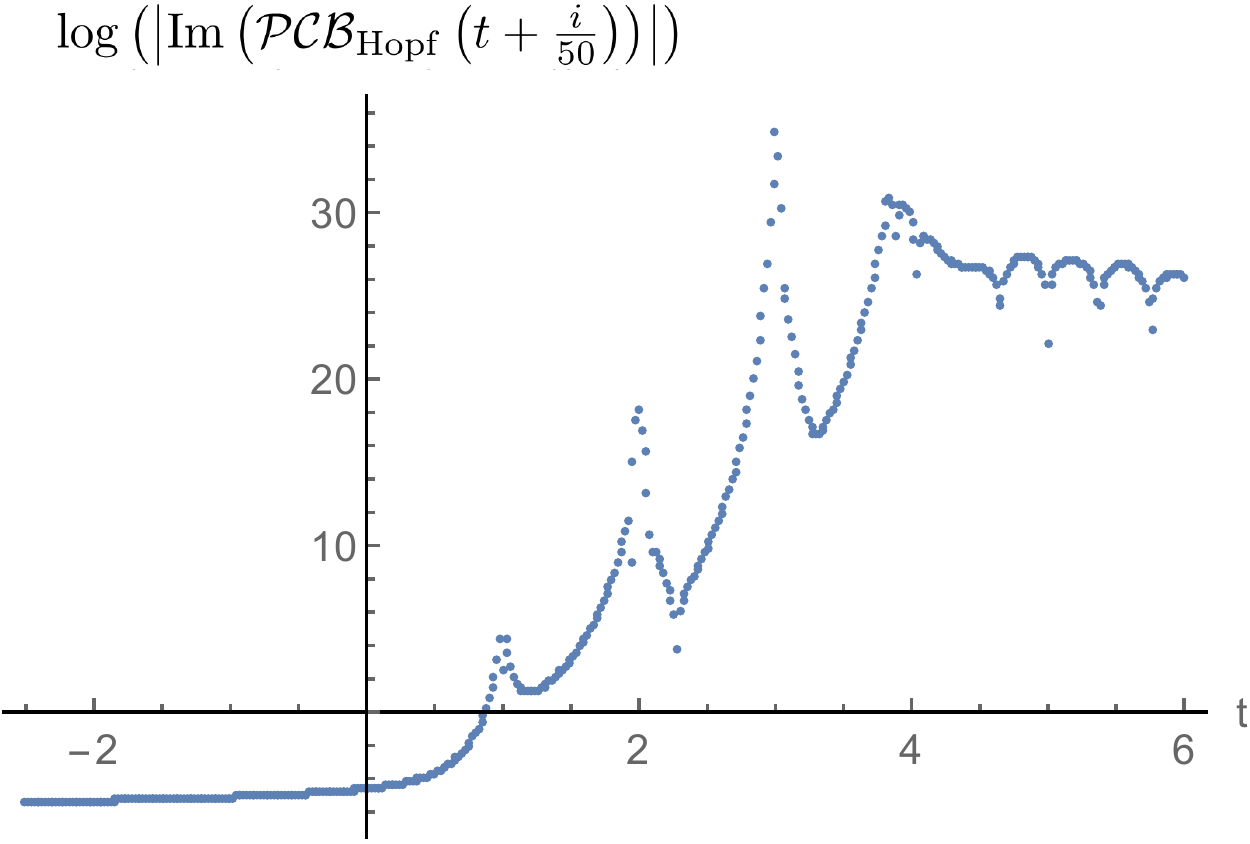}
 \caption{Plot of $\log \left(\left| {\rm Im} \left(\mathcal{PCB}_{\rm Hopf}\left(t+\frac{\rm i}{50}\right)\right)\right| \right)$ for the Pad\'e--conformal--Borel extrapolation. The logarithm is plotted in order to show all the singularities on the same figure.
 The peaks at $t=+1,+2,+3,+4$ indicate the location of a branch point in the Borel $t$ plane.}
 \label{fig:PertPCBLogPlot}
\end{figure}

This structure of higher Borel singularities can also be seen after mapping the Pad\'e approximant back to the Borel $t$ plane (step 4 of the ${\mathcal{PCB}}$ algorithm above). Figure~\ref{fig:PertPCBLogPlot} shows a log plot of the imaginary part of the Borel transform just above the real $t$ axis, revealing singularities at $t=+1,+2,+3,+4$ in the Borel $t$ plane, beyond the leading singularity at $t=+1$.
This Pad\'e--conformal--Borel analysis confirms our physical expectation that there should be higher Borel singularities at integer multiples of the leading one at $t=+1$, corresponding to a ``multi-instanton expansion''. Furthermore, these higher Borel singularities all appear to be branch point singularities.
These {\it numerical} results, and further extensions thereof, are derived {\it analytically} using the method of trans-series in Section \ref{sec:trans}.

To conclude this Borel analysis section, we comment further on the comparison with the chain approximation in (\ref{eq:chain}), which also produces a divergent formal perturbative expansion. Adopting the same scaling as in the Hopf expansion in (\ref{eq:g}), in terms of the variable $x=-a/3$, the corresponding formal chain approximation expansion is
\begin{align}
G_{\rm chain}(x)&:=\gamma_{\rm chain}(-3 x) \nonumber
\\
&\sim 3\sum_{n=1}^\infty \frac{\Gamma(n)}{2^n} \bigg(1-\frac{2}{2^n}+\frac{1}{3^n}\bigg) x^n \nonumber
\\
&=-24\int_0^\infty \frac{{\rm d}t}{(t-2)(t-4)(t-6)} {\rm e}^{-t/x}.
\label{eq:g-chain}
\end{align}
The final expression here has the form of a Borel integral,\footnote{{This formal integral expression is of course to be understood in the Borel--\'Ecalle sense~\cite{costin2008asymptotics,Dorigoni:2014hea,ecalle1981fonctions,mitschi2016divergent,sauzin}.}} with Borel singularities consisting of just three simple poles: at $t=+2$, $t=+4$ and $t=+6$. The singularities are poles rather than branch points, and there are only three of them, not an infinite ``multi-instanton'' tower of integer-spaced singularities. Also note that the leading Borel singularity for the chain expansion is at {\it twice} the distance from the origin compared to the leading singularity of the Borel transform for the Hopf expansion. This corresponds directly to the fact that the chain expansion is less divergent than the Hopf expansion: compare (\ref{eq:an}) and (\ref{eq:chaina}), or (\ref{eq:pert-growth1}) and (\ref{eq:chain-an}). However, despite these significant differences between the chain and Hopf approximations, we note one interesting similarity: in both cases the first three Borel singularity locations are in the relative proportions $1:2:3$. In the next section a trans-series analysis reveals analytically a similar feature of the Borel transform of the Hopf expansion, and explains the physical significance of this fact.

\section[Trans-series analysis of the Hopf-algebraic Dyson--Schwinger equation]
{Trans-series analysis of the Hopf-algebraic \\Dyson--Schwinger equation}
\label{sec:trans}

The perturbative solution for $G(x)$ in (\ref{eq:g-exp})--(\ref{eq:g-exp2}) cannot be the general solution of the Dyson--Schwinger equation (\ref{eq:ode2}), because the general solution of this third order equation involves 3~boundary condition parameters. The perturbative expansion in (\ref{eq:g-exp}) has no boundary condition parameters. The resolution of this mis-match is to generalize the divergent formal perturbative series in (\ref{eq:g-exp})--(\ref{eq:g-exp2}) to a {\it trans-series expansion}, which may involve exponentially small non-perturbative corrections and also possibly logarithmic terms. The trans-series involves three trans-series parameters, which we denote as $\sigma_1$, $\sigma_2$, $\sigma_3$, and which characterize the boundary conditions associated with a three-parameter family of solutions, each of which shares the same divergent formal power series perturbative expansion in (\ref{eq:g-exp}). The full structure of the trans-series can be extracted from the differential equation (\ref{eq:ode2}), and exhibits a rich network of resurgence relations connecting different non-perturbative sectors~\cite{costin-imrn,costin1998,costin2008asymptotics}.

\subsection{Identifying the \texorpdfstring{``seed''}{"seed"} exponential terms: the linearized equation}
\label{sec:seed0}

The first step in constructing the trans-series from the differential equation (\ref{eq:ode2}) is to identify the basic exponential ``building blocks'', or ``seeds'', of the trans-series. To do this, we linearize~(\ref{eq:ode2}) with the ansatz form~\cite{costin-imrn,costin1998,costin2008asymptotics}
\begin{gather}
G(x)\sim G^{\rm pert}(x)+G^{\rm non\text{-}pert}(x), \qquad x\to 0^+.
\label{eq:ansatz1}
\end{gather}
Here $G^{\rm non\text{-}pert}(x)$ is exponentially small, beyond all orders in perturbation theory as $x\to 0^+$, also with an accompanying prefactor power of $x$:
\begin{gather}
G^{\rm non\text{-}pert}(x)\sim x^\beta {\rm e}^{-\lambda/x}(1+O(x)), \qquad x\to 0^+.
\label{eq:ansatz2}
\end{gather}
We substitute the ansatz (\ref{eq:ansatz1})--(\ref{eq:ansatz2}) into the Dyson--Schwinger equation (\ref{eq:ode2})
and {\it linearize} the equation for $G^{\rm non\text{-}pert}(x)$. This results in the following third order linear and homogeneous equation for $G_{\rm non\text{-}pert}(x)$:
\begin{gather}
8 x^3 G_{\rm pert}(x)^3 G_{\rm non\text{-}pert}'''(x)
\nonumber
\\[.3ex] \qquad
{}+\Big(32 x^3 G_{\rm pert}(x)^2 G_{\rm pert}'(x)+8 x^2 G_{\rm pert}(x)^3-24 x^2 G_{\rm pert}(x)^2\Big) G_{\rm non\text{-}pert}''(x)
\nonumber
\\[.3ex] \qquad
{} + \Big(24 x^3 G_{\rm pert}(x) G_{\rm pert}'(x)^2+8 x^2 G_{\rm pert}(x)^2 G_{\rm pert}'(x) -48 x^2 G_{\rm pert}(x) G_{\rm pert}'(x)
\nonumber
\\[.3ex] \qquad\phantom{ +\Big(}
{} +32 x^3 G_{\rm pert}(x)^2 G_{\rm pert}''(x)\!+\!4 x G_{\rm pert}(x)^3\!+\!12 x G_{\rm pert}(x)^2\!+\!22 x G_{\rm pert}(x)\Big)G_{\rm non\text{-}pert}'(x)
 \nonumber
 \\[.3ex] \qquad
{} +\Big(64 x^3 G_{\rm pert}(x) G_{\rm pert}'(x) G_{\rm pert}''(x)+8 x^3 G_{\rm pert}'(x)^3+8 x^2 G_{\rm pert}(x) G_{\rm pert}'(x)^2
 \nonumber
 \\[.3ex] \qquad\phantom{ +\Big(}
{} -24 x^2 G_{\rm pert}'(x)^2+24 x^2 G_{\rm pert}(x)^2 G_{\rm pert}''(x)-48 x^2 G_{\rm pert}(x) G_{\rm pert}''(x)
 \nonumber
 \\[.3ex] \qquad\phantom{ +\Big(}
{}+24 x^3 G_{\rm pert}(x)^2 G_{\rm pert}'''(x)
 +12 x G_{\rm pert}(x)^2 G_{\rm pert}'(x)+24 x G_{\rm pert}(x) G_{\rm pert}'(x)
 \nonumber
 \\[.3ex] \qquad \phantom{ +\Big(}
{} +22 x G_{\rm pert}'(x)\!-\!4 G_{\rm pert}(x)^3\!
 -\!18 G_{\rm pert}(x)^2\!-22 G_{\rm pert}(x)\!-\!6\Big)G_{\rm non\text{-}pert}(x) =0.
 \label{eq:mess}
\end{gather}

This linear equation for $G^{\rm non\text{-}pert}(x)$ determines three possible pairs of solutions (since it is a third order equation) for the parameters $\beta$ and $\lambda$ in the non-perturbative ansatz (\ref{eq:ansatz2}).
It is convenient to express these in a vector notation:
\begin{gather}
{\vec \lambda}=(1, 2, 3) , \qquad {\vec \beta}=\bigg({-}\frac{23}{12}, +\frac{1}{6}, -\frac{11}{4}\bigg).
\label{eq:vector}
\end{gather}
The existence of three different values of the parameter $\lambda$ corresponds to the fact that for the perturbative Hopf expansion (\ref{eq:g}) there are three Borel singularities at different locations: at $t=1, 2, 3$. This explains analytically the existence of Borel singularities at $t=1, 2, 3$, which were found numerically by the conformal mapping analysis of the Borel transform in the previous section. Recall Figures \ref{fig:pcb-z-poles} and \ref{fig:PertPCBLogPlot}.
Note also the similarity to, and difference from, the Borel singularities of the chain approximation expansion, which appear at $t=2, 4, 6$ (in the same normalization: see equation~(\ref{eq:g-chain})).

The linearized equation {(\ref{eq:mess})} for $G^{\rm non\text{-}pert}(x)$ also generates the corresponding subleading power-law factor, $x^\beta$, characterized by the $\beta$ parameter in (\ref{eq:ansatz2}). The fact that the three values of $\beta$ in (\ref{eq:vector}) are all different and are all non-integer implies that the corresponding Borel singularities at $t=1, 2, 3$ are branch point singularities, each with a different exponent
\cite{costin-imrn,costin1998,costin2008asymptotics}.
We denote the solutions of the linearized equation as:
\begin{gather}
G_{\vec{k}}^{\text{non-pert, linearized}}(x) := \big(\vec{k}\cdot\vec{\sigma}\big) x^{{\vec k}\cdot {\vec \beta}} {\rm e}^{-({\vec k}\cdot {\vec \lambda})/x} {\mathcal F}_{\vec k}(x).
\label{eq:g-lin}
\end{gather}
Here we label the three different solutions (\ref{eq:g-lin}) to the third order linearized equation by the three basis unit vectors
\begin{eqnarray}
\vec{k}=(1, 0, 0), \qquad \vec{k}=(0, 1, 0), \qquad \vec{k}=(0, 0, 1).
\label{eq:k-vec}
\end{eqnarray}
Thus, $\beta_1=(1, 0, 0)\cdot\vec{\beta}=-\frac{23}{12}$, and $\lambda_1=(1, 0, 0)\cdot \vec{\lambda}=1$, and so on. Notice that each of the three ``seed'' solutions in (\ref{eq:g-lin}) has a free multiplicative parameter, $\sigma_1$, $\sigma_2$, $\sigma_3$, since each solves the {third order {\it homogeneous linear} equation (\ref{eq:mess}).}
For these seed solutions of the linearized equation, the final factor in (\ref{eq:g-lin}), denoted as ${\mathcal F}_{\vec k}(x)$, is a formal fluctuation series (also labeled by ${\vec k}$):\footnote{We index the coefficients $a_n^{\vec{k}}$ beginning with $n=1$ to match the indexing convention for the perturbative coefficients $A_n$ in (\ref{eq:an}).}
\begin{gather}
{\mathcal F}_{\vec k}(x) := \sum_{n=1}^\infty a^{{\vec k}}_{n} x^{n-1}.
\label{eq:fk}
\end{gather}

The coefficients $a^{{\vec k}}_{n}$ in (\ref{eq:fk}) are generated recursively by simple substitution of the ansatz~(\ref{eq:g-lin}) into the differential equation
(\ref{eq:ode2})
\begin{gather}
 \big\{a^{(1,0,0)}_n\big\} = \bigg\{{-}1,\frac{97}{48},\frac{53917}{13824},\frac{3026443}{221184},\frac{32035763261}{382205952} ,\dots \bigg\},
 \label{eq:a100}
 \\
 \big\{a^{(0,1,0)}_n\big\} =\bigg\{{-}1,\frac{151}{24},-\frac{63727}{3456},\frac{7112963}{82944},-\frac{7975908763}{23887872} ,\dots \bigg\},
 \label{eq:a010}
 \\
 \big\{a^{(0,0,1)}_n\big\} = \bigg\{{-}1,\frac{227}{48},\frac{1399}{4608},\frac{814211}{73728},\frac{3444654437}{42467328},\dots\bigg\}.
 \label{eq:a001}
\end{gather}

\begin{itemize}\itemsep=0pt
\item
We observe that the $a^{(1, 0, 0)}_n$ coefficients in (\ref{eq:a100}) coincide with {(up to an overall minus sign)} the coefficients of the subleading corrections to the large-order growth of the perturbative coefficients $A_n$ in (\ref{eq:pert-growth2})--(\ref{eq:pert-growth3}). This can be confirmed to very high subleading order. This is a clear example of the generic low-order/large-order resurgent behavior connecting the large order growth of the perturbative series to the low orders of the fluctuations around the first instanton term~\cite{BerryHowls}.
\item
We also observe that the pre-factor exponent, $(1, 0, 0)\cdot \vec{\beta}=-\frac{23}{12}$, associated with the leading exponential term, ${\rm e}^{-((1,0,0)\cdot {\vec \lambda})/x}={\rm e}^{-1/x}$, coincides with the offset of the argument of the leading factorial growth of the perturbative $A_n$ coefficients in (\ref{eq:pert-growth1}). This is also an example of generic behavior relating a formal perturbative series with the leading exponential correction.
\end{itemize}

We record the leading large order behavior of the expansion coefficients in (\ref{eq:a100})--(\ref{eq:a001}).
We~generated 500 coefficients (\ref{eq:a100})--(\ref{eq:a001}) (all rational numbers) for each of the fluctuation expansions in (\ref{eq:fk}). Each series is factorially divergent. The leading large-order growth of the expansion coefficients is\footnote{Notice that the coefficients $a^{(0,1,0)}_n$ alternate in sign at low order, but eventually settle down to have the same sign.}
 \begin{gather}
 a^{(1,0,0)}_n = 4 S_1 \Gamma \bigg(n+\frac{23}{12}\bigg) \big(1+\bigO\big(n^{-1}\big)\big),
 \qquad n\to \infty,
 \label{eq:a100-large}
 \\
 a^{(0,1,0)}_n = 4 S_1 \Gamma \bigg(n+\frac{23}{12}\bigg) \big(1+\bigO\big(n^{-5/6}\big)\big),
 \qquad n\to \infty,
 \label{eq:a010-large}
 \\
 a^{(0,0,1)}_n = \frac{20}{3} S_1 \Gamma \bigg(n+\frac{23}{12}\bigg) \big(1+\bigO\big(n^{-1}\big)\big), \qquad n\to \infty.
 \label{eq:a001-large}
 \end{gather}
Note that the factorial growth factor is the same for each series, and also agrees with that of the perturbative series; recall (\ref{eq:pert-growth1}). Also note that the overall Stokes constants are expressed as simple rational multiples of the Stokes constant $S_1$ for the perturbative series: recall (\ref{eq:pert-growth1}) and (\ref{eq:S1}). Thus, at this leading order level, no new independent Stokes constant is generated. Corrections to these {\it leading} large order growth expressions in (\ref{eq:a100-large})--(\ref{eq:a001-large}) are discussed in~Section~\ref{sec:logn}.

\subsection{Beyond the linearized equation: resonant trans-series and logarithms}

The three exponential terms, ${\rm e}^{-\lambda_1/x}$, ${\rm e}^{-\lambda_2/x}$ and ${\rm e}^{-\lambda_3/x}$, are just the ``seed'' non-perturbative terms coming from the {\it linearized} equation {(\ref{eq:mess}) for $G^{\rm non\text{-}pert}(x)$. But the full} Dyson--Schwinger equation is nonlinear, so each of these three exponential terms will re-appear in all integer powers of the seed term, generating a trans-series expansion that includes both perturbative and non-perturbative terms to all orders. For a {\it generic} {\it non-resonant} trans-series~\cite{costin-imrn,costin1998,costin2008asymptotics}, in which there are no integer relations between the different $\lambda_j$ values in (\ref{eq:vector}), the trans-series would be generated as a three dimensional infinite sum generating all powers of the basic seed factors, $x^{{\vec k}\cdot {\vec \beta}} {\rm e}^{-({\vec k}\cdot {\vec \lambda})/x}$, each being further multiplied by a formal fluctuation series. The terms in such a~{\it non-resonant} trans-series would all have the form indicated in (\ref{eq:g-lin}), but now labeled by vectors~$\vec{k}$ being integer linear combinations of the basis vectors in (\ref{eq:k-vec})~\cite{costin-imrn,costin1998,costin2008asymptotics}.

However, our case here is {\it resonant}. This is because the Dyson--Schwinger equation (\ref{eq:ode2}) has the special {\it resonant} property that the three different possible values for the exponent $\lambda$ in (\ref{eq:ansatz2}) satisfy integer relations:
\begin{gather*}
\lambda_1 : \lambda_2 :\lambda_3=1 : 2 : 3.
\end{gather*}
This resonant property has profound implications for the structure of the trans-series beyond the leading exponential order, leading to an even richer structure.\footnote{A familiar and illustrative example is the Painlev\'e I equation, a second-order nonlinear equation which has two resonant values $\lambda=\pm 1$, suitably normalized~\cite{Aniceto:2011nu,Garoufalidis:2010ya}. Here the resonant structure is quite different.}
In a resonant case, the exponent coefficient $({\vec k}\cdot {\vec \lambda})$ in (\ref{eq:g-lin}) may take the same value for different integer-valued vectors~${\vec k}$. For example, ${\rm e}^{-2/x}$ appears through one power of the seed term with $\lambda_2=2$, but it also appears via the square of the seed term with $\lambda_1=1$.
Therefore, when we grade the solution by its exponential order,
${\rm e}^{-(\text{integer})/x}$, a given order can have
contributions from different ${\vec k}$ vectors, and when they mix there can also appear logarithmic terms in the solution to the Dyson--Schwinger equation. An analogous feature is familiar for linear ODEs, but for a nonlinear ODE there will appear higher powers of logarithms as we go higher in the exponential order of the trans-series (i.e., to higher ``instanton'' order).

To characterize this resonant structure we write the trans-series expansion in an exponentially graded form
\begin{gather}
G(x)\sim G_{(0)}(x) \!+ {\rm e}^{-1/x} G_{(1)}(x) \!+ {\rm e}^{-2/x} G_{(2)}(x) \!+ {\rm e}^{-3/x} G_{(3)}(x) \!+ {\rm e}^{-4/x} G_{(4)}(x)+\cdots.
\label{eq:graded}
\end{gather}
Here $G_{(0)}(x)\equiv G^{\text{pert}}(x)$, is the formal perturbative series in (\ref{eq:g})--(\ref{eq:g-exp}). By construction, each~$G_{(m)}(x)$, for $m\geq 1$, satisfies a third order {\it linear} differential equation. For the first exponential term, $G_{(1)}(x)$, this equation is both {\it linear} and {\it homogeneous}, and the solution is:
\begin{gather}
G_{(1)}(x) = \sigma_1 x^{-23/12} {\mathcal F}_{(1, 0, 0)}(x).
\label{eq:g1}
\end{gather}
This corresponds to the first of the ``seed'' solutions (\ref{eq:g-lin}) from the previous section, with the exponential factor separated out as in (\ref{eq:graded}).
Up to this first exponential order there are no logarithmic terms.

At second exponential order, and beyond, the equation for $G_{(m)}(x)$, with $m\geq 2$, is linear but {\it inhomogeneous}. We thus need a particular solution in addition to the homogeneous solution.
For $m=2$ the inhomogeneity comes from a term involving $\big(G_{(1)}(x)\big)^2$.
This means that the inhomogeneity vanishes if $G_{(1)}(x)$ vanishes. This is the case if the first trans-series parameter~$\sigma_1$ is chosen to vanish. In this case $G_{(2)}(x)$ satisfies a linear {\it homogeneous} equation, and there exists a~solution, which is multiplied by a new constant parameter, the second trans-series parameter~$\sigma_2$:
\begin{eqnarray}
G^{\rm homogeneous}_{(2)}(x)=\sigma_2 x^{1/6} \mathcal F_{(0, 1, 0)}(x), \qquad \text{with}\quad \sigma_1=0.
\label{eq:g2f}
\end{eqnarray}
Noting that $(0, 1, 0)\cdot\vec{\beta}=\frac{1}{6}$, we recognize this as corresponding to the second ``seed'' solution in~(\ref{eq:g-lin}), once again with the exponential factor separated out.
However, if $\sigma_1\neq 0$, the full solution to the {\it inhomogeneous} equation for $G_{(2)}(x)$ requires also a particular solution.
To generate this solution
we substitute a Frobenius ansatz:
\begin{gather*}
G_{(2)}(x) \sim x^{\delta}\Bigg[ \sum_n\Delta_n x^n + \bigg(\sum_n \mu_n x^n \bigg) \log(x) \Bigg].
\end{gather*}
The power parameter $\delta$, and the expansion coefficients $\Delta_n$ and $\mu_n$ are determined recursively by the linear inhomogeneous equation for $G_{(2)}(x)$.
This leads to the solution
\begin{gather}
G_{(2)}(x)\sim \bigg(\frac{1}{x^{23/12}}\bigg)^2 \bigg[{-}2\sigma_1^2 \frac{1}{x} \mathcal F_{(2, 0, 0)}(x)+ x^4 \bigg(\sigma_1^2\frac{21265 }{2304}\log(x) + \sigma_2 \bigg) \mathcal{F}_{(0,1,0)}(x)\bigg],
 \label{eq:g2full}
 \end{gather}
 where we find a new fluctuation series $\mathcal F_{(2, 0, 0)}(x)$ \big(the labeling notation refers to the two powers of $\sigma_1$ multiplying this term: $\sigma_1^2=\sigma_1^2\sigma_2^0\sigma_3^0:=\vec{\sigma}^{(2,0,0)}$\big):
 \begin{gather}
\mathcal F_{(2, 0, 0)}(x) \sim 1-\frac{49}{12}x-\frac{13235}{3456}x^2-\frac{43049}{3456}x^3 \nonumber
 -\frac{2496477497}{23887872}x^4 -0\cdot x^5
 \\ \hphantom{\mathcal F_{(2, 0, 0)}(x) \sim}
 {}-\frac{3315185066507813}{247669456896}x^6 - \cdots.
\label{eq:f200}
\end{gather}
Several comments are in order concerning the structure of the expression for $G_{(2)}(x)$ in (\ref{eq:g2full})--(\ref{eq:f200}):
\begin{enumerate}\itemsep=0pt
\item
The overall rational-exponent prefactor in $G_{(2)}(x)$ is naturally expressed in terms of the square of the corresponding rational-exponent prefactor for $G_{(1)}(x)$ in (\ref{eq:g1}).
 \item
 Noting that $\big(\frac{1}{x^{23/12}}\big)^2 x^4=x^{1/6}$, we see that
 when $\sigma_1=0$ we recover the second ``seed'' solution $G^{\rm homogeneous}_{(2)}(x)$ in (\ref{eq:g2f}), with fluctuation series
 $\mathcal{F}_{(0,1,0)}(x)$, and multiplied by its new trans-series parameter $\sigma_2\equiv (0,1,0)\cdot\vec{\sigma}$.

 \item
 When $\sigma_1\neq 0$ there appears a $\log(x)$ term, proportional to $\sigma_1^2$, and multiplied by the same ``seed'' fluctuation series $\mathcal{F}_{(0,1,0)}(x)$ mentioned in the previous item, and with a specific fixed coefficient: $\frac{21265 }{2304} x^4$.
 \item
 When $\sigma_1\neq 0$ there is also a new series, once again proportional to $\sigma_1^2$, denoted as $\mathcal F_{(2, 0, 0)}(x)$, and whose first terms are listed in (\ref{eq:f200}).
 \item
The coefficient of $x^5$ in $\mathcal F_{(2, 0, 0)}(x)$ vanishes.
Note that because the value of $\sigma_2$ is not fixed and behaves like an integration constant, the fact that this coefficient vanishes is not a~coincidence, but a convenient {\it choice}. In fact, we could have fixed this coefficient to any value while appropriately changing higher order terms of $\mathcal F_{(2, 0, 0)}(x)$.

\end{enumerate}

At third exponential order in the graded trans-series (\ref{eq:graded}), the function $G_{(3)}(x)$ satisfies a linear and inhomogeneous equation, and the inhomogeneity involves both $(G_{(1)}(x))^3$ and $G_{(1)}(x) G_{(2)}(x)$. Thus, setting $\sigma_1=0$, we obtain a {\it homogeneous} linear equation for $G_{(3)}(x)$, with solution:
\begin{gather}
G^{\rm homogeneous}_{(3)}(x)=\sigma_3 x^{-11/4} \mathcal F_{(0, 0, 1)}(x), \qquad \text{with}\quad \sigma_1=0.
\label{eq:g3f}
\end{gather}
We recognize this as corresponding to the third ``seed'' solution in (\ref{eq:g-lin}), with $\vec{k}=(0,0,1)$ and with the third independent factor $\sigma_3$ (and once again with the exponential factor separated out).
But when $\sigma_1\neq 0$, resonant logarithmic terms appear again.
For $G_{(3)}(x)$ we find an exp\-res\-sion involving two new series, denoted $\mathcal F_{(3, 0, 0)}(x)$ and $\mathcal F_{(1, 1, 0)}(x)$, multiplied by $\sigma_1^3$ and~$\sigma_1\sigma_2$, respectively:
\begin{gather}
G_{(3)}(x)\sim \bigg(\frac{1}{x^{23/12}}\bigg)^3 \bigg[\sigma_1^3 \bigg({-}\frac{6}{x^2}\bigg)
\mathcal F_{(3, 0, 0)}(x)+ x^3 \bigg({-}5\sigma_1^3 \frac{21265 }{2304}\log(x) + \sigma_3 \bigg) \mathcal{F}_{(0, 0, 1)}(x) \nonumber
\\ \hphantom{G_{(3)}(x)\sim\bigg(\frac{1}{x^{23/12}}\bigg)^3 \bigg[}
{}+12 x^4\bigg(\sigma_1^3\frac{21265}{2304}\log(x) + \sigma_1\sigma_2\bigg) \mathcal{F}_{(1,1, 0)}(x) \bigg].
 \label{eq:g3full}
 \end{gather}
 The two new series, $\mathcal F_{(3, 0, 0)}(x)$ and $\mathcal F_{(1, 1, 0)}(x)$, appearing in (\ref{eq:g3full}) have initial terms:
 \begin{gather}
 \mathcal F_{(3, 0, 0)}(x) \sim 1-\frac{103 x}{16}+\frac{8821 x^2}{4608}-\frac{454379 x^3}{221184}-\frac{1344528799 x^4}{14155776}+0\cdot x^5
 \nonumber
 \\ \hphantom{ \mathcal F_{(3, 0, 0)}(x) \sim }
 {}-\frac{9242013290874467x^6}{587068342272} +\cdots,
 \label{eq:f300}
 \\
 \mathcal F_{(1, 1, 0)}(x) \sim 1-\frac{131 x}{16}+\frac{153997 x^2}{4608}-\frac{12555605 x^3}{73728}+\frac{484910403 x^4}{524288}-\frac{4580515441493x^5}{679477248}
 \nonumber
 \\ \hphantom{ \mathcal F_{(1, 1, 0)}(x) \sim}
{} +\frac{28651912194246539 x^6}{587068342272}-\frac{30305985730060738841 x^7}{65751654334464} + \cdots.
 \label{eq:f110}
 \end{gather}
We comment on the structure of the expression for $G_{(3)}(x)$ in (\ref{eq:g3full})--(\ref{eq:f110}):
\begin{enumerate}\itemsep=0pt
\item
The overall rational-exponent prefactor in $G_{(3)}(x)$ is naturally expressed in terms of the cube of the corresponding rational-exponent prefactor for $G_{(1)}(x)$ in (\ref{eq:g1}).
 \item
 Noting that $\big(\frac{1}{x^{23/12}}\big)^3 x^3=x^{-11/4}$, we see that
 when $\sigma_1=0$ we recover the third ``seed'' solution $G^{\rm homogeneous}_{(3)}(x)$ in (\ref{eq:g3f}), with fluctuation series
 $\mathcal{F}_{(0,0,1)}(x)$, and multiplied by its new trans-series parameter $\sigma_3\equiv (0,0,1)\cdot\vec{\sigma}$.

 \item
 When $\sigma_1\neq 0$ there appears a $\log(x)$ term, proportional to $\sigma_1^3$. Remarkably, this logarithmic term is multiplied by a rational-exponent factor $x^{-11/4}=x^{(0,0,1)\cdot\vec{\lambda}}$, and by a fluctuation series that naturally splits into two pieces. One piece coincides with the fluctuation series $\mathcal{F}_{(0,0,1)}(x)$ appearing in the homogeneous solution $G^{\rm homogeneous}_{(3)}(x)$ in (\ref{eq:g3f}), while the other piece is $x$ times the fluctuation factor $\mathcal{F}_{(1,1,0)}(x)$ which multiplies the $\sigma_1\sigma_2\equiv \vec{\sigma}^{(1,1,0)}$ factor in the full solution (\ref{eq:g3full}).

 \item
 When $\sigma_1\neq 0$ there is also a new series, proportional to $\sigma_1^3$, denoted as $\mathcal F_{(3, 0, 0)}(x)$, and whose first terms are listed in (\ref{eq:f300}). Note that the coefficient of $x^5$ in $\mathcal F_{(3, 0, 0)}(x)$ vanishes, just like for $\mathcal F_{(2, 0, 0)}(x)$ in (\ref{eq:f200}).

\end{enumerate}

Proceeding to higher exponential orders of the graded trans-series in (\ref{eq:graded}), one finds that there are no new homogeneous solutions, even after setting $\sigma_1=\sigma_2=\sigma_3=0$. This is because the three independent homogeneous solutions were generated by the linearized equation discussed in Section \ref{sec:seed0}.
Hence no new independent trans-series parameters are generated, which is consistent with the interpretation of $(\sigma_1,\sigma_2,\sigma_3)$ as the three boundary condition parameters of the third order Dyson--Schwinger equation (\ref{eq:ode2}).
The inhomogeneity in the linear equation for the 4th order graded term $G_{(4)}(x)$ involves various combinations of lower-order functions: $(i)$~4~factors involving $G_{(1)}(x)$; $(ii)$ 2 factors involving $G_{(1)}(x)$ and 1 factor involving $G_{(2)}(x)$; $(iii)$ 1 factor involving $G_{(1)}(x)$ and 1 factor involving $G_{(3)}(x)$; $(iv)$ 2 factors involving $G_{(2)}(x)$. Since one power of $\log(x)$ appears in $G_{(2)}(x)$, we find that the solution for $G_{(4)}(x)$ involves the first appearance of a $\log^2(x)$ term. This can be confirmed from the ODE by direct substitution.
 In general, continuing to all orders in the exponential grading we see that eventually all powers of $\log(x)$ are generated. However, at any given exponential order only a finite number of powers of $\log(x)$ appear. This trans-series structure is a direct consequence of the {\it resonant} character of the Dyson--Schwinger equation (\ref{eq:ode2}). Thus, the full trans-series solution involves all powers of the three basic {\it trans-monomial} elements: $x$, ${\rm e}^{-1/x}$, and $\log(x)$,\footnote{The fact that only the three basic {\it trans-monomial} elements, $x$, ${\rm e}^{-1/x}$, and $\log(x)$, suffice to construct the trans-series solution of the ODE follows from~\cite{costin-imrn,costin1998,costin2008asymptotics}.} with $\log(x)$ terms appearing only at the second exponential order and beyond. The physical significance of this fact is discussed in the conclusions.

\subsection{Logarithmic behavior in large-order growth}\label{sec:logn}

The appearance of $\log(x)$ terms in the trans-series starting at the second exponential order of the graded trans-series (\ref{eq:graded}) is reflected in the appearance of $\log(n)$ corrections appearing in the subleading large-order growth of the coefficients of the fluctuation series beyond the original perturbative series $G_{(0)}(x)\equiv G_{\rm pert}(x)$.\footnote{Analogous logarithmic large-order growth is seen in the Painlev\'e I equation~\cite{Aniceto:2011nu,Garoufalidis:2010ya}.} To illustrate this phenomenon, consider the coefficients~$a_n^{(1,0,0)}$ of the fluctuation series ${\mathcal F}_{(1,0,0)}(x)$, which appears in the first exponential term~$G_{(1)}(x)$ in~(\ref{eq:g1}). The first few coefficients were already listed in (\ref{eq:a100}) and their leading factorial growth was shown in (\ref{eq:a100-large}). To probe the subleading corrections to this {\it leading} factorial growth, it is instructive to begin with a conventional resurgent ansatz involving power-law subleading corrections:
\begin{gather}
 \Gamma\bigg(n+\frac{23}{12}\bigg)\Bigg[1+\sum_{k=1}^\infty \frac{b_k}{\prod_{l=1}^k \big(n+\frac{23}{12} -l\big)} \Bigg].
\label{eq:form}
\end{gather}
Here the $b_k$ are expected to be rational numbers related to coefficients of fluctuation expansions higher in the trans-series.
We will soon see that this ansatz (\ref{eq:form}) is in fact insufficient, but remarkably it is extremely precise for the first 4 subleading power-law corrections.

\begin{figure}[htb!]
\centering
\includegraphics[width=100mm]{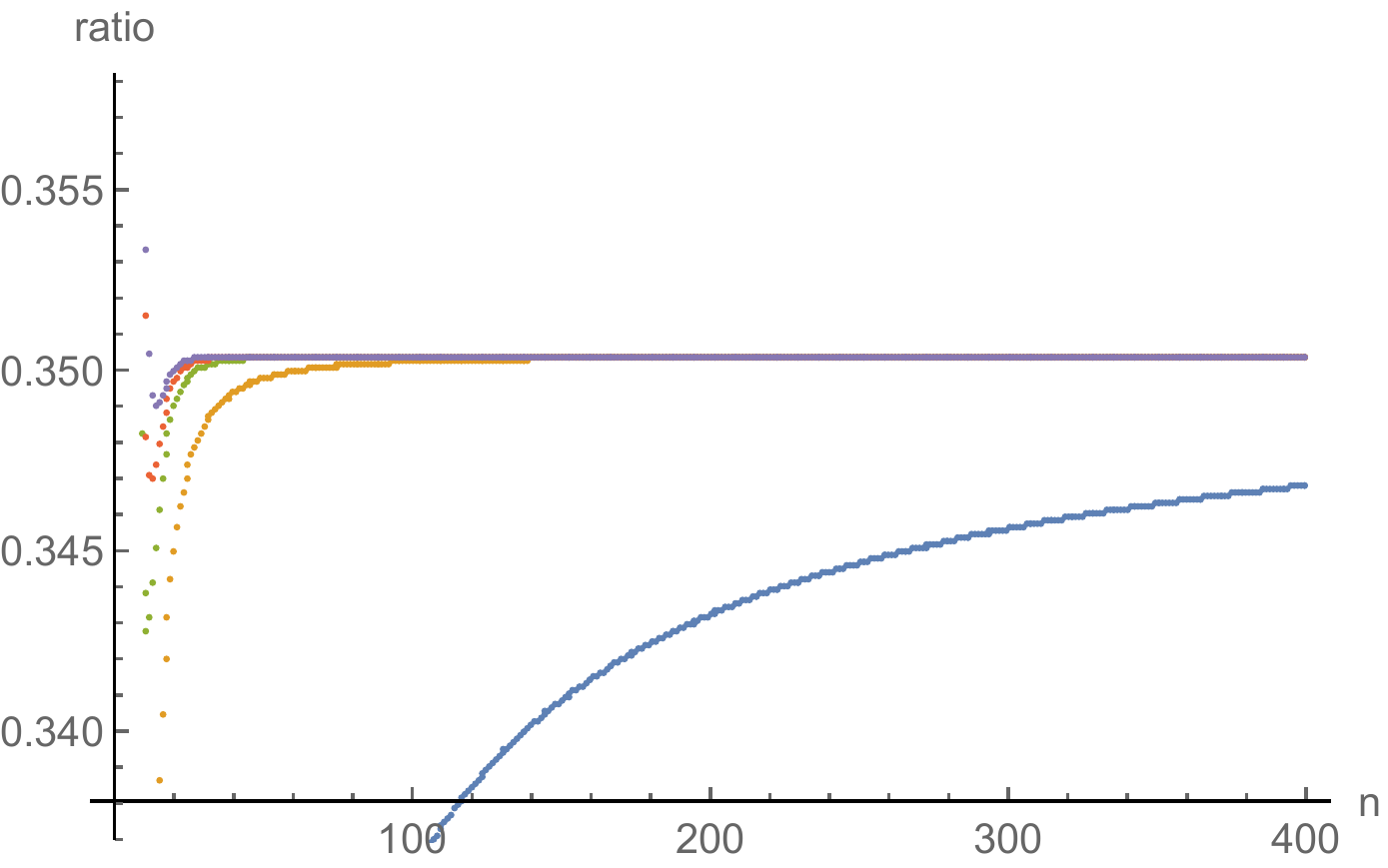}
 \caption{Plot of the ratio of the coefficients $a^{(1,0,0)}_n$ to the leading growth $\Gamma\left(n+\frac{23}{12}\right)$ in (\ref{eq:a100-large}) (blue curve), and then subsequent curves include the further subleading large $n$ corrections shown in (\ref{eq:g2-resurgence3}). The large $n$ values tend to $4 S_1=0.3503822\dots$, in terms of the Stokes constant of the large order growth of the original perturbative coefficients in (\ref{eq:pert-growth1})--(\ref{eq:S1}).}
 \label{fig:a100-sub}
\end{figure}

Using 500 terms of the $a^{(1, 0, 0)}_n$ coefficients, we use high-order Richardson extrapolation methods to identify the first few such rational coefficients:
 \begin{gather}
 a^{(1, 0, 0)}_n \sim 4 S_1\Gamma\bigg(n+\frac{23}{12}\bigg)\Bigg[1-\frac{\frac{49}{12}}{\big(n+\frac{11}{12}\big)}
 -\frac{\frac{13235}{3456}}{\big(n-\frac{1}{12}\big) \big(n+\frac{11}{12}\big)}
 -\frac{\frac{43049}{3456}}{\big(n-\frac{13}{12}\big)\big(n-\frac{1}{12}\big) \big(n+\frac{11}{12}\big)} \nonumber
 \\ \hphantom{ a^{(1, 0, 0)}_n \sim 4 S_1\Gamma\bigg(n+\frac{23}{12}\bigg)\bigg[}
{} -\frac{\frac{2496477497}{23887872}}{ \big(n-\frac{25}{12}\big) \big(n-\frac{13}{12}\big) \big(n-\frac{1}{12}\big) \big(n+\frac{11}{12}\big)}
 -\cdots \Bigg].
 \label{eq:g2-resurgence3}
 \end{gather}
 Figure~\ref{fig:a100-sub} illustrates the extra precision gained by including these subleading corrections; compare with the analogous plot in Figure~\ref{fig:subleading-an} for the perturbative series coefficients. The precision is extremely good for the first four subleading corrections displayed in (\ref{eq:g2-resurgence3}).

Notice that the first coefficients of the subleading correction terms for large-order growth of the coefficients $a^{(1,0,0)}_n$, as shown in (\ref{eq:g2-resurgence3}), coincide with the low order coefficients of the fluctuation series $\mathcal F_{(2, 0, 0)}(x)$ in (\ref{eq:f200}). Thus, superficially this looks like a generic large-order/low-order resurgence relation connecting the large orders of the series coefficients entering the ${\rm e}^{-1/x}G_{(1)}(x)$ term in the graded trans-series (\ref{eq:graded}) and the ${\rm e}^{-2/x} G_{(2)}(x)$ term in (\ref{eq:graded}). However, this correspondence breaks down at the next subleading order. This can be traced to the fact that at this order in $G_{(2)}(x)$ we find a logarithmic term $x^5\log(x)$. Moreover, the coefficient of this logarithmic term is fixed in (\ref{eq:g2full}). This suggest that we extend the large-order growth ansatz in (\ref{eq:form}) to incorporate a $\log(n)$ growth factor:
 \begin{gather}
 a^{(1, 0, 0)}_n \sim 4 S_1 \Gamma\bigg(n+\frac{23}{12}\bigg)\Bigg[1-\frac{\frac{49}{12}}{\left(n+\frac{11}{12}\right)}
 -\frac{\frac{13235}{3456}}{\left(n-\frac{1}{12}\right) \left(n+\frac{11}{12}\right)}
 \nonumber
 \\[1ex]\hphantom {a^{(1, 0, 0)}_n \sim}
{} -\frac{\frac{43049}{3456}}{\left(n-\frac{13}{12}\right)\left(n-\frac{1}{12}\right) \left(n+\frac{11}{12}\right)}
 -\frac{\frac{2496477497}{23887872}}{ \left(n-\frac{25}{12}\right) \left(n-\frac{13}{12}\right) \left(n-\frac{1}{12}\right) \left(n+\frac{11}{12}\right)}
 \nonumber
 \\[1ex]\hphantom {a^{(1, 0, 0)}_n \sim}
{} +\frac{d-\frac{21265 }{4608} \log(n)}{\left(n-\frac{37}{12}\right) \left(n-\frac{25}{12}\right) \left(n-\frac{13}{12}\right) \left(n-\frac{1}{12}\right) \left(n+\frac{11}{12}\right)}
 +\cdots \Bigg].
 \label{eq:g2-resurgence4}
 \end{gather}
The logarithmic term in front of $\log(n)$ can be found using a modified Richardson acceleration method, as described in Appendix~\ref{sec:app}.
 The coefficient is found to be rational and in correspondence with the rational factor $\frac{21265 }{2304}$ multiplying the $\log(x)$ term in $G^{(2)}(x)$ in (\ref{eq:g2full}). The~constant coefficient $d$ in (\ref{eq:g2-resurgence4}) can similarly be determined to extremely high precision but does not appear to be rational:
 \begin{gather}
d=
-846.3135357762392334586752585470377294931556414335975009793245
\dots.
\label{eq:d}
\end{gather}

\section{Conclusions}

In this paper we have shown that the non-perturbative completion of the perturbative Hopf-algebraic Dyson--Schwinger equation (\ref{eq:ode2}) for the anomalous dimension of the massless $\phi^3$ scalar QFT in 6 dimensions leads to a trans-series of the following form:
\begin{gather}
G(x)\sim \sum_{n=0}^\infty \sum_{k=0}^\infty \sum_{l=0}^{L_k} c_{n, k, l} x^{n+1-k} \bigg(\frac{{\rm e}^{-1/x}}{x^{23/12}}\bigg)^k \big(\log(x)\big)^l.
\label{eq:general}
\end{gather}
We stress that this trans-series was derived directly from the Dyson--Schwinger equation (\ref{eq:ode2}), which was itself derived from the manifestly {\it perturbative} Hopf-algebraic renormalization for\-malism. No semi-classical non-perturbative argument was invoked; just a resurgent asymptotic analysis of the Dyson--Schwinger differential equation. Nevertheless, we observe that the re\-sulting trans-series has the structure of a multi-instanton expansion, typical of semi-classical computations in quantum mechanics and quantum field theory. Each multi-instanton term,~${\rm e}^{-k/x}$, is multiplied by a linear combination of formal asymptotic perturbative series in powers of~$x$, and polynomial factors in powers of logarithms of~$x$. The first logarithmic term arises at two-instanton order, and at a given instanton order there is a maximal power $L_k=[k/2]$ of $\log(x)$ factors. Remarkably, this structure matches the trans-series structure found using instanton-calculus methods in quantum mechanical models and also in certain quantum field theories~\cite{Dunne:2012ae,Dunne:2016nmc,Garoufalidis:2010ya, ZinnJustin:2002ru,ZinnJustin:2004ib,ZinnJustin:2004ib+}. We note that other infinitely-iterated approximations to the Dyson--Schwinger equations, such as the rainbow or chain approximation, do not lead to a~multi-instanton trans-series structure.

In semi-classical instanton-calculus computations the logarithmic terms arise from quasi-zero modes associated with bion solutions, which can be interpreted as molecules of instantons and anti-instantons~\cite{Dunne:2016nmc, ZinnJustin:2002ru}.
By contrast, our starting point was purely perturbative, with the non-perturbative features being recovered from a trans-series analysis of the Hopf-algebraic Dyson--Schwinger equation. The sum over instanton-like exponential terms arises from the generic integer-spaced repetition of Borel singularities in the solution of the nonlinear Dyson--Schwinger differential equation for the anomalous dimension, while the logarithmic terms in (\ref{eq:general}) come from the special {\it resonant} structure of the nonlinear differential equation. This nonlinear equation is third order, so generically there would be three independent Borel singularities~\cite{costin-imrn,costin1998,costin2008asymptotics}, therefore associated with three independent ``instantons''. However, the Dyson--Schwinger equation has extra symmetry that results in the three Borel singularities being collinear, and also in the integer proportions $1: 2: 3$. This resonant form can be traced to the particular iterative structure of the Hopf-algebraic Dyson--Schwinger equations, and has interesting implications for the resurgent large-order/low-order relations. Resonance produces $\log(n)$ factors in the large-order growth of the coefficients of the fluctuations about the instanton factors, but absent in the large order growth of the original formal perturbative series. Similar $\log(n)$ behavior is seen in the large order growth of fluctuation coefficients at higher orders of a semiclassical instanton calculus computation.

While there is a long history of studying logarithmic terms in semiclassical approximations in
quantum mechanical and quantum field theoretics models, and in resonant nonlinear differential equations,
 it is much more rare to be able to identify such behavior in terms of the renormalized coupling, and especially in an asymptotically free QFT. Renormalons, which are also associated with iterative perturbative structures, have been studied recently using ideas from resurgence, in a wide variety of theories: see for example~\cite{Anber:2014sda,Dondi:2020qfj,Ishikawa:2019oga,Maiezza:2018pkk,Marino:2019eym,Marino:2019fvu, Marino:2020dgc}, and references therein.
 There are interesting analogies between the iterative Hopf Dyson--Schwinger formalism and renormalon and large $N$ computations in QFT, and these issues are currently under investigation. It would also be valuable to understand in detail the relation to a quite different Hopf approximation for the $\phi_6^3$ theory, which includes some aspects of vertex corrections~\cite{Bellon:2020qlx}. Ultimately the goal is to understand more deeply how the non-perturbative trans-series structure fits naturally within the underlying perturbative Hopf algebra structure of QFT renormalization.

The Dyson--Schwinger ODE \eqref{eq:ode2} could also be analysed using \'Ecalle's alien differential calculus~\cite{ecalle1981fonctions,mitschi2016divergent}. For the 4 dimensional Yukawa theory analysed in~\cite{Borinsky:2020vae}, an alien calculus approach led to a closed form expression for the all orders trans-series for the anomalous dimension. In this Yukawa model the trans-series contained no logarithmic terms, so the alien derivative formalism of~\cite{borinsky2018generating} was well adapted to derive the closed form solution. For the 6 dimensional $\phi^3$ model, the appearance of resonant logarithmic terms produces a structure that lies beyond the simplified alien calculus framework of~\cite{borinsky2018generating}. This motivates a generalization of~\cite{borinsky2018generating} that uses more aspects of \'Ecalle's full theory of resurgence, a project beyond the scope of this article, and left for future work. Such a formalism would have many potential applications. It could enable
 a systematic evaluation of the large order behaviour at all orders, as was done in the Yukawa model~\cite{Borinsky:2020vae}. An alien calculus treatment should also prove to all orders the correspondence between the trans-series fluctation coefficients and the large order behaviour of the expansion coefficients. A~rigorous treatment of the $\log$ term asymptotic behaviour would be especially interesting as there are intriguing parallels to resonance phenomena in quantum mechanics, where the appearance of the $\log$ terms can be associated with the inherent stability properties of the underlying quantum mechanical system.

\appendix
\section[Modified Richardson method for probing logarithmic large-order growth]
{Modified Richardson method for probing logarithmic \\large-order growth}
\label{sec:app}

In this appendix we explain an effective general method to obtain high-precision numerical results for the coefficients of high-order growth terms involving $\log(n)$ and powers of $\log(n)$. For applications to $\log(n)$ growth see, for example,~\cite{Garoufalidis:2010ya,ZinnJustin:1979db}. We first summarize the conventional Richardson extrapolation method~\cite{bender-book} in a form that makes the generalization simple to formulate, and also simple to implement.

As input data, we have a sequence of numbers $f_n$, whose asymptotic behaviour is to be determined experimentally. Even though it is not technically necessary, it increases the efficacy of the method dramatically if the numbers $f_n$ are given as explicit rational numbers. If the numbers are not represented as rational numbers, they have to be available up to very high precision for the method to give reliable results.
To apply the original version of Richardson extrapolation, we need to assume that the numbers $f_n$ have the asymptotic behaviour,
\begin{gather}
 \label{eq:normalasymp}
 f_n \sim \sum_{k = 0}^{K} a_k n^{-k} + \bigO\big(n^{-K-1}\big) \qquad\text{as}\quad n \rightarrow \infty,
\end{gather}
for some arbitrary $K \geq 0$ with certain (not necessarily known) coefficients $a_k$. This condition is satisfied by the subleading correction factor in the conventional resurgent large-order growth ansatz (\ref{eq:form}) truncated to order $K$, and matches, for example, the subleading large-order growth in (\ref{eq:pert-growth2}).
In determining the coefficients of such subleading corrections, the technical task is to extract effectively the coefficients $a_n$ in \eqref{eq:normalasymp}, starting with $a_0$ and then proceeding to higher coefficients. The precision should be high enough to be able to recognize rational values of the coefficients, whose rationality can then be tested by probing further corrections.

It follows immediately from \eqref{eq:normalasymp} that $f_n \sim a_0 + \bigO\big(n^{-1}\big)$. Therefore, if we naively use the value of $f_n$ for some large $n$ as an estimate for $a_n$, the error will be of the order $\frac{1}{n}$. Due to this slow rate of convergence, we would need a huge number of coefficients $f_n$ to get a sufficiently accurate estimate. Richardson extrapolation provides a more efficient way to determine the value of $a_0$ at higher precision with fewer input terms.

Let $\Delta_n$ be the (forward) difference operator, i.e., for any sequence $f_n$
\begin{gather}
\label{eq:forward}
\Delta_n f_n = f_{n+1} - f_n.
\end{gather}
Clearly, the operator $\Delta_n$ is linear and can be iterated. We treat it as a derivative operator that acts on everything on its right. For instance,
\begin{gather*}
\Delta_n^2 f_n = \Delta_n (\Delta_n f_n) = \Delta_n f_{n+1} - \Delta_n f_n = f_{n+2} - 2 f_{n+1} + f_n.
\end{gather*}
With this operator the $K$-th order Richardson extrapolation of a sequence $f_n$ can be defined as,
\begin{gather}
 \label{eq:normalrich}
 R_K [ f_n ] := \frac{1}{K!} \Delta_n^K n^K f_n.
\end{gather}
This form of Richardson extrapolation is well adapted to straightforward implementation beca\-use the difference operator is a natural computer operation.
The key observation concerning extrapolation is the fact that if the sequence $f_n$ has the asymptotic form in \eqref{eq:normalasymp}, then
\begin{gather}
 \label{eq:normalrichasymp}
 R_K [ f_n ] = \frac{1}{K!} \Delta_n^K n^K f_n \sim a_0 + \bigO\big(n^{-K-1}\big) \qquad\text{as}\quad n \rightarrow \infty.
\end{gather}
Thus, the Richardson extrapolated sequence $R_K[f_n]$ converges much more rapidly, with order $\bigO\big(n^{-K-1}\big)$. If $K$ is chosen sufficiently large, the Richardson extrapolation \eqref{eq:normalrich} can therefore be used to estimate $a_0$ to very high accuracy.

To verify \eqref{eq:normalrichasymp}, observe that by \eqref{eq:normalasymp} the sequence $n^N f_n$ has the asymptotic form,
\begin{align}
\label{eq:nn}
n^K f_n \sim p(n) + \sum_{k = 1}^\infty a_{K+k} n^{-k} \qquad\text{as}\quad n \rightarrow \infty,
\end{align}
where $p(n) = a_0 n^K + \bigO\big(n^{K-1}\big)$ is a polynomial of order $K$. In order to prove~\eqref{eq:normalrichasymp} from~\eqref{eq:nn}, we need to use the definition \eqref{eq:normalrich} and also the following elementary properties of the $\Delta_n$-operator:
\begin{gather}
\Delta_n^K n^K = K!, \qquad \Delta_n^K n^k =0 \qquad\text{for all}\quad 0 \leq k < K,\nonumber
\\
\Delta_n^K n^{-k} \sim \bigO\big(n^{-K-k}\big)\qquad \text{for all}\quad k > 0 \qquad\text{as}\quad n\rightarrow \infty.\label{eq:richid1}
\end{gather}

Having formulated conventional Richardson extrapolation in terms of the forward difference operator \eqref{eq:forward}, it is now straightforward to construct generalizations that can be applied to sequences with a more general asymptotic behaviour than that in \eqref{eq:normalasymp}. For example, suppose a sequence $g_n$ has the following asymptotic behaviour with a subleading $\log n$ contribution:
\begin{gather*}
 \label{eq:logasymp}
 g_n \sim \sum_{k = 0}^{K} a_k n^{-k} + \log n \sum_{k = 1}^{K} b_k n^{-k} + \bigO\big(n^{-K-1} \log n\big) \qquad\text{as}\quad n \rightarrow \infty.
\end{gather*}
Naive application of the Richardson extrapolation operator does not lead to an accelerated convergence, due to the extra $\log$-terms.

To derive the variant of Richardson extrapolation that can also deal with this more general asymptotic behaviour, we first
evaluate some identities for the difference operator applied to $\log$-terms. We have,
\begin{align*}
\Delta_n^K n^{k-1} \log n &=
\Delta_n^{K-1} \bigg( \big((n+1)^{k-1} - n^{k-1}\big)\log n + (n+1)^{k-1} \log\bigg(1+\frac{1}{n}\bigg) \bigg)
\\
&\sim \sum_{\ell=K-k+1}^K \widetilde c_\ell n^{-\ell} + \bigO\big(n^{-K-1}\big) \qquad \text{for all}\quad 1 \leq k \leq K \quad\text{as}\quad n \rightarrow \infty,
\end{align*}
for some coefficients $\widetilde c_\ell$ whose specific values are not important for this discussion, and where we used the expansion $\log\big(1+\frac{1}{n}\big) = -\sum_{k=1}^{\infty} \frac{(-1)^{k}}{k} n^{-k}$. Similarly, we obtain
\[
\Delta_n^K n^{-k} \log n \sim \bigO\big(n^{-k-K} \log n\big)\qquad \text{for all}\quad k > 0\quad
\quad \text{as}\quad n\rightarrow \infty.
\]
Using these observations together with the previous ones in \eqref{eq:richid1}, we find that the application of the Richardson extrapolation operator $\Delta_n^K n^K$ on the sequence $g_n$ leads to a sequence with the following asymptotic behaviour:
\begin{gather*}
 R_K[g_n] = \frac{1}{K!} \Delta_n^K n^K g_n \sim a_0 + \sum_{k=1}^{K} c_k n^{-k} + \bigO\big(n^{-K-1} \log n\big) \qquad \text{as}\quad n \rightarrow \infty,
\end{gather*}
with some coefficients $c_k$. The $\log$ terms are suppressed now, so we can apply the normal Richardson $R_K$ operator in \eqref{eq:normalrich} again, in order to get rid of the subleading non-$\log$-terms with equation~\eqref{eq:normalrichasymp}, and obtain a rapidly converging sequence,
\begin{gather*}
 R_K[R_K[g_n]] = \frac{1}{K!^2} \Delta_n^K n^K \Delta_n^K n^K g_n \sim a_0 + \bigO\big(n^{-K-1} \log n\big) \qquad \text{as}\quad n \rightarrow \infty.
\end{gather*}
This can be used for efficient extraction of the $a_0$ term. Note that this double Richardson extrapolation operator is equivalent to a difference operator of order $2K$.

Analogously, we can derive related difference operators that produce rapidly converging sequ\-ences for other kinds of asymptotic behaviours. If, for instance, the sequence $h_n$ is expected to behave asymptotically as follows with a leading $\log$-term,
\begin{gather*}
 h_n \sim \log n \sum_{k = 0}^{K} a_k n^{-k} + \sum_{k = 0}^{K} b_k n^{-k} + \bigO\big(n^{-K-1} \log n\big) \qquad\text{as}\quad n \rightarrow \infty,
\end{gather*}
and we are again interested in the leading $a_0$ coefficient, then the following $2K+1$-th order difference operator applied to $h_n$ leads to the desired result:
\begin{gather*}
\frac{1}{K!^2} \Delta_n^K n^{K+1} \Delta_n^{K+1} n^K h_n \sim a_0 + \bigO\big(n^{-K-1} \log n\big) \qquad \text{as}\quad n \rightarrow \infty.
\end{gather*}
The proof works analogously to the previous derivations.
This is the operator that we used to verify the expected asymptotic behaviour in \eqref{eq:g2-resurgence4}, obtaining sufficient precision to clearly identify the rational coefficient $\frac{21265 }{4608}$, which is associated with the coefficient of the $\log(x)$ term in $G^{(2)}(x)$ in (\ref{eq:g2full}).
Difference operators that can deal with sequences involving higher power $\log^\ell$-terms can be easily developed in a similar fashion.

\subsection*{Acknowledgements}
This material is based upon work supported by the U.S.~Department of Energy, Office of Science, Office of High Energy Physics under Award Number DE-SC0010339 (GD, MM) and by the NWO Vidi grant 680-47-551 ``Decoding Singularities of Feynman graphs'' (MB). This work was begun during visits by the first two authors to Humboldt University in 2018 and 2019, and at the Les Houches Summer School in 2018, and we thank these institutions for hospitality. We are grateful to Marc Bellon, David Broadhurst, Ovidiu Costin, John Gracey, Dirk Kreimer, Enrico Russo and Karen Yeats for discussions. We also want to thank David Broadhurst for helping with the estimation of the constant $d$ in equation~\eqref{eq:d}, and for pointing out some typos in a~previous version.


\pdfbookmark[1]{References}{ref}
\LastPageEnding


\begin{thebibliography}{99}
\footnotesize\itemsep=0pt

\bibitem{alvarez_howls_silverstone}
\'Alvarez G., Howls C.J., Silverstone H.J., Anharmonic oscillator discontinuity
 formulae up to second-exponentially-small order, \href{https://doi.org/10.1088/0305-4470/35/18/302}{\textit{J.~Phys.~A: Math.
 Gen.}} \textbf{35} (2002), 4003--4016.

\bibitem{Anber:2014sda}
Anber M.M., Sulejmanpasic T., The renormalon diagram in gauge theories on
 {$\mathbb R^3\times \mathbb S^1$}, \href{https://doi.org/10.1007/JHEP01(2015)139}{\textit{J.~High Energy Phys.}}
 \textbf{2015} (2015), no.~1, 139, 34~pages, \href{https://arxiv.org/abs/1410.0121}{arXiv:1410.0121}.

\bibitem{Aniceto:2018bis}
Aniceto I., Ba\c{s}ar G., Schiappa R., A primer on resurgent transseries and
 their asymptotics, \href{https://doi.org/10.1016/j.physrep.2019.02.003}{\textit{Phys. Rep.}} \textbf{809} (2019), 1--135,
 \href{https://arxiv.org/abs/1802.10441}{arXiv:1802.10441}.

\bibitem{Aniceto:2011nu}
Aniceto I., Schiappa R., Vonk M., The resurgence of instantons in string
 theory, \href{https://doi.org/10.4310/CNTP.2012.v6.n2.a3}{\textit{Commun. Number Theory Phys.}} \textbf{6} (2012), 339--496,
 \href{https://arxiv.org/abs/1106.5922}{arXiv:1106.5922}.

\bibitem{Bellon:2010sf}
Bellon M.P., An efficient method for the solution of {S}chwinger--{D}yson
 equations for propagators, \href{https://doi.org/10.1007/s11005-010-0415-3}{\textit{Lett. Math. Phys.}} \textbf{94} (2010),
 77--86, \href{https://arxiv.org/abs/1005.0196}{arXiv:1005.0196}.

\bibitem{Bellon:2016mje}
Bellon M.P., Clavier P.J., Alien calculus and a {S}chwinger--{D}yson equation:
 two-point function with a nonperturbative mass scale, \href{https://doi.org/10.1007/s11005-017-1016-1}{\textit{Lett. Math.
 Phys.}} \textbf{108} (2018), 391--412, \href{https://arxiv.org/abs/1612.07813}{arXiv:1612.07813}.

\bibitem{Bellon:2020uzi}
Bellon M.P., Russo E.I., Resurgent analysis of {W}ard--{S}chwinger--{D}yson
 equations, \href{https://doi.org/10.3842/SIGMA.2021.075}{\textit{SIGMA}} \textbf{17} (2021), 075, 18~pages,
 \href{https://arxiv.org/abs/2011.13822}{arXiv:2011.13822}.

\bibitem{Bellon:2020qlx}
Bellon M.P., Russo E.I., Ward--{S}chwinger--{D}yson equations in {$\phi^3_6$}
 quantum field theory, \href{https://doi.org/10.1007/s11005-021-01377-2}{\textit{Lett. Math. Phys.}} \textbf{111} (2021), 42,
 31~pages, \href{https://arxiv.org/abs/2007.15675}{arXiv:2007.15675}.

\bibitem{Bellon:2008zz}
Bellon M.P., Schaposnik F.A., Renormalization group functions for the
 {W}ess--{Z}umino model: up to 200 loops through {H}opf algebras,
 \href{https://doi.org/10.1016/j.nuclphysb.2008.02.005}{\textit{Nuclear Phys.~B}} \textbf{800} (2008), 517--526, \href{https://arxiv.org/abs/0801.0727}{arXiv:0801.0727}.

\bibitem{bender-book}
Bender C.M., Orszag S.A., Advanced mathematical methods for scientists and
 engineers.~{I}. {A}symptotic methods and perturbation theory,
 \href{https://doi.org/10.1007/978-1-4757-3069-2}{Springer-Verlag}, New York, 1999.

\bibitem{Benedetti:2020iku}
Benedetti D., Delporte N., Remarks on a melonic field theory with cubic
 interactions, \href{https://doi.org/10.1007/JHEP04(2021)197}{\textit{J.~High Energy Phys.}} \textbf{2021} (2021), no.~4, 197,
 30~pages, \href{https://arxiv.org/abs/2012.12238}{arXiv:2012.12238}.

\bibitem{BerryHowls}
Berry M.V., Howls C.J., Hyperasymptotics for integrals with saddles,
 \href{https://doi.org/10.1098/rspa.1991.0119}{\textit{Proc. Roy. Soc. London Ser.~A}} \textbf{434} (1991), 657--675.

\bibitem{Bersini:2019axn}
Bersini J., Maiezza A., Vasquez J.C., Resurgence of the renormalization group
 equation, \href{https://doi.org/10.1016/j.aop.2020.168126}{\textit{Ann. Physics}} \textbf{415} (2020), 168126, 15~pages,
 \href{https://arxiv.org/abs/1910.14507}{arXiv:1910.14507}.

\bibitem{Bogner:2017xhp}
Bogner C., Borowka S., Hahn T., Heinrich G., Jones S.P., Kerner M., von
 Manteuffel A., Michel M., Panzer E., Papara V., Loopedia, a database for loop
 integrals, \href{https://doi.org/10.1016/j.cpc.2017.12.017}{\textit{Comput. Phys. Commun.}} \textbf{225} (2018), 1--9,
 \href{https://arxiv.org/abs/1709.01266}{arXiv:1709.01266}.

\bibitem{borinsky2018generating}
Borinsky M., Generating asymptotics for factorially divergent sequences,
 \href{https://doi.org/10.37236/5999}{\textit{Electron.~J. Combin.}} \textbf{25} (2018), 4.1, 32~pages,
 \href{https://arxiv.org/abs/1603.01236}{arXiv:1603.01236}.

\bibitem{borinsky2018graphs}
Borinsky M., Graphs in perturbation theory: algebraic structure and
 asymptotics, \textit{Springer Theses}, \href{https://doi.org/10.1007/978-3-030-03541-9}{Springer}, Cham, 2018.

\bibitem{Borinsky:2020rqs}
Borinsky M., Tropical {M}onte {C}arlo quadrature for {F}eynman integrals,
 \href{https://arxiv.org/abs/2008.12310}{arXiv:2008.12310}.

\bibitem{Borinsky:2020vae}
Borinsky M., Dunne G.V., Non-perturbative completion of {H}opf-algebraic
 {D}yson--{S}chwinger equations, \href{https://doi.org/10.1016/j.nuclphysb.2020.115096}{\textit{Nuc\-lear Phys.~B}} \textbf{957} (2020),
 115096, 17~pages, \href{https://arxiv.org/abs/2005.04265}{arXiv:2005.04265}.

\bibitem{Borinsky:2021jdb}
Borinsky M., Gracey J.A., Kompaniets M.V., Schnetz O., Five-loop
 renormalization of {$\phi^3$} theory with applications to the {L}ee--{Y}ang
 edge singularity and percolation theory, \href{https://doi.org/10.1103/physrevd.103.116024}{\textit{Phys. Rev.~D}} \textbf{103}
 (2021), 116024, 35~pages, \href{https://arxiv.org/abs/2103.16224}{arXiv:2103.16224}.

\bibitem{BoSc21graphical}
Borinsky M., Schnetz O., Graphical functions in even dimensions,
 \href{https://arxiv.org/abs/2105.05015}{arXiv:2105.05015}.

\bibitem{Brezin:1976vw}
Brezin E., Le~Guillou J.C., Zinn-Justin J., Perturbation theory at large order.
 {I}.~The $\phi^{2N}$ interaction, \href{https://doi.org/10.1103/PhysRevD.15.1544}{\textit{Phys. Rev.~D}} \textbf{15} (1977),
 1544--1557.

\bibitem{Broadhurst:1999ys}
Broadhurst D.J., Kreimer D., Combinatoric explosion of renormalization tamed by
 {H}opf algebra: 30-loop {P}ad\'e--{B}orel resummation, \href{https://doi.org/10.1016/S0370-2693(00)00051-4}{\textit{Phys. Lett.~B}}
 \textbf{475} (2000), 63--70, \href{https://arxiv.org/abs/hep-th/9912093}{arXiv:hep-th/9912093}.

\bibitem{Broadhurst:2000dq}
Broadhurst D.J., Kreimer D., Exact solutions of {D}yson--{S}chwinger equations
 for iterated one loop integrals and propagator coupling duality,
 \href{https://doi.org/10.1016/S0550-3213(01)00071-2}{\textit{Nuclear Phys.~B}} \textbf{600} (2001), 403--422,
 \href{https://arxiv.org/abs/hep-th/0012146}{arXiv:hep-th/0012146}.

\bibitem{caliceti}
Caliceti E., Meyer-Hermann M., Ribeca P., Surzhykov A., Jentschura U.D., From
 useful algorithms for slowly convergent series to physical predictions based
 on divergent perturbative expansions, \href{https://doi.org/10.1016/j.physrep.2007.03.003}{\textit{Phys. Rep.}} \textbf{446}
 (2007), 1--96, \href{https://arxiv.org/abs/0707.1596}{arXiv:0707.1596}.

\bibitem{caprini}
Caprini I., Fischer J., Abbas G., Ananthanarayan B., Perturbative expansions in
 {QCD} improved by conformal mappings of the {B}orel plane, in Perturbation
 Theory: Advances in Research and Applications, Nova Science Publishers, Inc.,
 Hauppauge, 2018, 211--254, \href{https://arxiv.org/abs/1711.04445}{arXiv:1711.04445}.

\bibitem{Cardy:1975fz}
Cardy J.L., High-energy behavior in $\phi^3$ theory in six-dimensions,
 \href{https://doi.org/10.1016/0550-3213(75)90518-0}{\textit{Nuclear Phys.~B}} \textbf{93} (1975), 525--546.

\bibitem{Clavier:2019sph}
Clavier P.J., Borel--\'Ecalle resummation of a two-point function, \href{https://doi.org/10.1007/s00023-021-01057-w}{\textit{Ann.
 Henri Poincar\'e}} \textbf{22} (2021), 2103--2136, \href{https://arxiv.org/abs/1912.03237}{arXiv:1912.03237}.

\bibitem{Connes:1999zw}
Connes A., Kreimer D., Renormalization in quantum field theory and the
 {R}iemann--{H}ilbert problem, \href{https://doi.org/10.1088/1126-6708/1999/09/024}{\textit{J.~High Energy Phys.}} \textbf{1999}
 (1999), no.~9, 024, 8~pages, \href{https://arxiv.org/abs/hep-th/9909126}{arXiv:hep-th/9909126}.

\bibitem{Connes:2000fe}
Connes A., Kreimer D., Renormalization in quantum field theory and the
 {R}iemann--{H}ilbert problem. {II}.~{T}he {$\beta$}-function, diffeomorphisms
 and the renormalization group, \href{https://doi.org/10.1007/PL00005547}{\textit{Comm. Math. Phys.}} \textbf{216}
 (2001), 215--241, \href{https://arxiv.org/abs/hep-th/0003188}{arXiv:hep-th/0003188}.

\bibitem{Cornwall:1995dr}
Cornwall J.M., Morris D.A., Toy models of nonperturbative asymptotic freedom in
 $\phi^3$ in six-dimensions, \href{https://doi.org/10.1103/PhysRevD.52.6074}{\textit{Phys. Rev.~D}} \textbf{52} (1995),
 6074--6086, \href{https://arxiv.org/abs/hep-ph/9506293}{arXiv:hep-ph/9506293}.

\bibitem{costin-imrn}
Costin O., Exponential asymptotics, transseries, and generalized {B}orel
 summation for analytic, nonlinear, rank-one systems of ordinary differential
 equations, \href{https://doi.org/10.1155/S1073792895000286}{\textit{Int. Math. Res. Not.}} \textbf{1995} (1995), 377--417,
 \href{https://arxiv.org/abs/math.CA/0608414}{arXiv:math.CA/0608414}.

\bibitem{costin1998}
Costin O., On {B}orel summation and {S}tokes phenomena for rank-{$1$} nonlinear
 systems of ordinary differential equations, \href{https://doi.org/10.1215/S0012-7094-98-09311-5}{\textit{Duke Math.~J.}}
 \textbf{93} (1998), 289--344, \href{https://arxiv.org/abs/math.CA/0608408}{arXiv:math.CA/0608408}.

\bibitem{costin2008asymptotics}
Costin O., Asymptotics and {B}orel summability, \textit{Chapman \& Hall/CRC
 Monographs and Surveys in Pure and Applied Mathematics}, Vol.~141, CRC Press,
 Boca Raton, FL, 2009.

\bibitem{Costin:2020hwg}
Costin O., Dunne G.V., Physical resurgent extrapolation, \href{https://doi.org/10.1016/j.physletb.2020.135627}{\textit{Phys. Lett.~B}}
 \textbf{808} (2020), 135627, 7~pages, \href{https://arxiv.org/abs/2003.07451}{arXiv:2003.07451}.

\bibitem{Costin:2020pcj}
Costin O., Dunne G.V., Uniformization and constructive analytic continuation of
 {T}aylor series, \href{https://arxiv.org/abs/2009.01962}{arXiv:2009.01962}.

\bibitem{Courtiel:2019dnq}
Courtiel J., Yeats K., Next-to{$^k$} leading log expansions by chord diagrams,
 \href{https://doi.org/10.1007/s00220-020-03691-7}{\textit{Comm. Math. Phys.}} \textbf{377} (2020), 469--501,
 \href{https://arxiv.org/abs/1906.05139}{arXiv:1906.05139}.

\bibitem{damburg}
Damburg R.J., Propin R.K., Graffi S., Grecchi V., Harrell E.M.,
 \v{C}\'{\i}\v{z}ek J., Paldus J., Silverstone H.J., $1/R$~expansion for
 $H_2^+$: analyticity, summability, asymptotics, and calculation of
 exponentially small terms, \href{https://doi.org/10.1103/PhysRevLett.52.1112}{\textit{Phys. Rev. Lett.}} \textbf{52} (1984),
 1112--1115.

\bibitem{deAlcantaraBonfim:1980pe}
de~Alcantara~Bonfim O.F., Kirkham J.E., McKane A.J., Critical exponents to
 order $\epsilon^3$ for $\phi^3$ models of critical phenomena in six
 $\epsilon$-dimension, \href{https://doi.org/10.1088/0305-4470/13/7/006}{\textit{J.~Phys.~A: Math. Gen.}} \textbf{13} (1980),
 L247--L251, {E}rratum,
 \href{https://doi.org/10.1088/0305-4470/13/12/529}{\textit{J.~Phys.~A: Math.
 Gen.}} \textbf{13} (1980), 3785--3785.

\bibitem{deAlcantaraBonfim:1981sy}
de~Alcantara~Bonfim O.F., Kirkham J.E., McKane A.J., Critical exponents for the
 percolation problem and the {Y}ang--Lee edge singularity, \href{https://doi.org/10.1088/0305-4470/14/9/034}{\textit{J.~Phys.~A:
 Math. Gen.}} \textbf{13} (1981), 2391--2413.

\bibitem{ddp}
Delabaere E., Dillinger H., Pham F., Exact semiclassical expansions for
 one-dimensional quantum oscillators, \href{https://doi.org/10.1063/1.532206}{\textit{J.~Math. Phys.}} \textbf{38}
 (1997), 6126--6184.

\bibitem{ddp+}
Delabaere E., Pham F., Resurgent methods in semi-classical asymptotics,
 \textit{Ann. Inst. H.~Poincar\'e Phys. Th\'eor.} \textbf{71} (1999), 1--94.

\bibitem{Delbourgo:1996nw}
Delbourgo R., Elliott D., McAnally D.S., Dimensional renormalization in
 $\phi^{3}$ theory: ladders and rainbows, \href{https://doi.org/10.1103/PhysRevD.55.5230}{\textit{Phys. Rev.~D}} \textbf{55}
 (1997), 5230--5233, \href{https://arxiv.org/abs/hep-th/9611150}{arXiv:hep-th/9611150}.

\bibitem{Dondi:2020qfj}
Dondi N.A., Dunne G.V., Reichert M., Sannino F., Towards the {QED} beta
 function and renormalons at {$1/N^2_f$} and {$1/N^3_f$}, \href{https://doi.org/10.1103/physrevd.102.035005}{\textit{Phys.
 Rev.~D}} \textbf{102} (2020), 035005, 15~pages, \href{https://arxiv.org/abs/2003.08397}{arXiv:2003.08397}.

\bibitem{Dorigoni:2014hea}
Dorigoni D., An introduction to resurgence, trans-series and alien calculus,
 \href{https://doi.org/10.1016/j.aop.2019.167914}{\textit{Ann. Physics}} \textbf{409} (2019), 167914, 38~pages,
 \href{https://arxiv.org/abs/1411.3585}{arXiv:1411.3585}.

\bibitem{Dunne:2012ae}
Dunne G.V., \"Unsal M., Resurgence and trans-series in quantum field theory:
 the {$\mathbb{CP}^{N-1}$} model, \href{https://doi.org/10.1007/JHEP11(2012)170}{\textit{J.~High Energy Phys.}} \textbf{2012}
 (2012), no.~11, 170, 85~pages, \href{https://arxiv.org/abs/1210.2423}{arXiv:1210.2423}.

\bibitem{Dunne:2013ada}
Dunne G.V., \"Unsal M., Generating nonperturbative physics from perturbation
 theory, \href{https://doi.org/10.1103/PhysRevD.89.041701}{\textit{Phys. Rev.~D}} \textbf{89} (2014), 041701, 5~pages,
 \href{https://arxiv.org/abs/1306.4405}{arXiv:1306.4405}.

\bibitem{Dunne:2016nmc}
Dunne G.V., \"Unsal M., New nonperturbative methods in quantum field theory:
 from large-{$N$} orbifold equivalence to bions and resurgence, \href{https://doi.org/10.1146/annurev-nucl-102115-044755}{\textit{Ann.
 Rev. Nuclear Part. Sci.}} \textbf{66} (2016), 245--272, \href{https://arxiv.org/abs/1601.03414}{arXiv:1601.03414}.

\bibitem{ecalle1981fonctions}
\'Ecalle J., Les fonctions r\'esurgentes, {T}ome~{I}, Les alg\`ebres de
 fonctions r\'esurgentes, \textit{Publications Math\'ematiques d'Orsay~81},
 Vol.~5, Universit\'e de Paris-Sud, Orsay, 1981.

\bibitem{Fei:2014yja}
Fei L., Giombi S., Klebanov I.R., Critical $O(N)$ models in $6-\epsilon$
 dimensions, \href{https://doi.org/10.1103/PhysRevD.90.025018}{\textit{Phys. Rev.~D}} \textbf{90} (2014), 025018, 19 pages,
 \href{https://arxiv.org/abs/1404.1094}{arXiv:1404.1094}.

\bibitem{Fisher:1978pf}
Fisher M.E., Yang--{L}ee edge singularity and $\phi^3$ field theory,
 \href{https://doi.org/10.1103/PhysRevLett.40.1610}{\textit{Phys. Rev. Lett.}} \textbf{40} (1978), 1610--1613.

\bibitem{Garoufalidis:2010ya}
Garoufalidis S., Its A., Kapaev A., Mari\~no M., Asymptotics of the instantons
 of {P}ainlev\'e~{I}, \href{https://doi.org/10.1093/imrn/rnr029}{\textit{Int. Math. Res. Not.}} \textbf{2012} (2012),
 561--606, \href{https://arxiv.org/abs/1002.3634}{arXiv:1002.3634}.

\bibitem{Giombi:2019upv}
Giombi S., Huang R., Klebanov I.R., Pufu S.S., Tarnopolsky G., {$O(N)$} model
 in {$4<d<6$}: instantons and complex {CFT}s, \href{https://doi.org/10.1103/physrevd.101.045013}{\textit{Phys. Rev.~D}}
 \textbf{101} (2020), 045013, 25~pages, \href{https://arxiv.org/abs/1910.02462}{arXiv:1910.02462}.

\bibitem{Gracey:2015tta}
Gracey J.A., Four loop renormalization of {$\phi^3$} theory in six dimensions,
 \href{https://doi.org/10.1103/PhysRevD.92.025012}{\textit{Phys. Rev.~D}} \textbf{92} (2015), 025012, 31~pages,
 \href{https://arxiv.org/abs/1506.03357}{arXiv:1506.03357}.

\bibitem{Gracey:2020baa}
Gracey J.A., Asymptotic freedom from the two-loop term of the {$\beta$}
 function in a cubic theory, \href{https://doi.org/10.1103/physrevd.101.125022}{\textit{Phys. Rev.~D}} \textbf{101} (2020),
 125022, 13~pages, \href{https://arxiv.org/abs/2004.14208}{arXiv:2004.14208}.

\bibitem{Gracey:2020tkk}
Gracey J.A., Ryttov T.A., Shrock R., Renormalization-group behavior of
 {$\phi^3$} theories in {$d = 6$} dimensions, \href{https://doi.org/10.1103/physrevd.102.045016}{\textit{Phys. Rev.~D}}
 \textbf{102} (2020), 045016, 9~pages, \href{https://arxiv.org/abs/2007.12234}{arXiv:2007.12234}.

\bibitem{Grinstein:2014xba}
Grinstein B., Stone D., Stergiou A., Zhong M., Challenge to the $a$ theorem in
 six dimensions, \href{https://doi.org/10.1103/PhysRevLett.113.231602}{\textit{Phys. Rev. Lett.}} \textbf{113} (2014), 231602,
 5~pages, \href{https://arxiv.org/abs/1406.3626}{arXiv:1406.3626}.

\bibitem{Houghton:1978dt}
Houghton A., Reeve J.S., Wallace D.J., High order behavior in $\phi^3$ field
 theories and the percolation problem, \href{https://doi.org/10.1103/PhysRevB.17.2956}{\textit{Phys. Rev.~B}} \textbf{15}
 (1977), 2956--2964.

\bibitem{Ishikawa:2019oga}
Ishikawa K., Morikawa O., Shibata K., Suzuki H., Takaura H., Renormalon
 structure in compactified spacetime, \href{https://doi.org/10.1093/ptep/ptz147}{\textit{Prog. Theor. Exp. Phys.}}
 (2020), 013B01, 14~pages, \href{https://arxiv.org/abs/1909.09579}{arXiv:1909.09579}.

\bibitem{jentschura2001improved}
Jentschura U.D., Soff G., Improved conformal mapping of the {B}orel plane,
 \href{https://doi.org/10.1088/0305-4470/34/7/316}{\textit{J.~Phys.~A: Math. Gen.}} \textbf{34} (2001), 1451--1457,
 \href{https://arxiv.org/abs/hep-ph/0006089}{arXiv:hep-ph/0006089}.

\bibitem{Kompaniets:2017yct}
Kompaniets M.V., Panzer E., Minimally subtracted six-loop renormalization of
 {$O(n)$}-symmetric {$\phi^4$} theory and critical exponents, \href{https://doi.org/10.1103/physrevd.96.036016}{\textit{Phys.
 Rev.~D}} \textbf{96} (2017), 036016, 26~pages, \href{https://arxiv.org/abs/1705.06483}{arXiv:1705.06483}.

\bibitem{Kreimer:1997dp}
Kreimer D., On the {H}opf algebra structure of perturbative quantum field
 theories, \href{https://doi.org/10.4310/ATMP.1998.v2.n2.a4}{\textit{Adv. Theor. Math. Phys.}} \textbf{2} (1998), 303--334,
 \href{https://arxiv.org/abs/q-alg/9707029}{arXiv:q-alg/9707029}.

\bibitem{Kreimer:2006ua}
Kreimer D., Yeats K., An \'etude in non-linear {D}yson--{S}chwinger equations,
 \href{https://doi.org/10.1016/j.nuclphysbps.2006.09.036}{\textit{Nuclear Phys.~B Proc. Suppl.}} \textbf{160} (2006), 116--121,
 \href{https://arxiv.org/abs/hep-th/0605096}{arXiv:hep-th/0605096}.

\bibitem{Kreimer:2006gm}
Kreimer D., Yeats K., Recursion and growth estimates in renormalizable quantum
 field theory, \href{https://doi.org/10.1007/s00220-008-0431-7}{\textit{Comm. Math. Phys.}} \textbf{279} (2008), 401--427,
 \href{https://arxiv.org/abs/hep-th/0612179}{arXiv:hep-th/0612179}.

\bibitem{Kruger:2019tas}
Kr\"uger O., Log expansions from combinatorial {D}yson--{S}chwinger equations,
 \href{https://doi.org/10.1007/s11005-020-01288-8}{\textit{Lett. Math. Phys.}} \textbf{110} (2020), 2175--2202,
 \href{https://arxiv.org/abs/1906.06131}{arXiv:1906.06131}.

\bibitem{Lapedes:1981tz}
Lapedes A., Mottola E., Complex path integrals and finite temperature,
 \href{https://doi.org/10.1016/0550-3213(82)90477-1}{\textit{Nuclear Phys.~B}} \textbf{203} (1982), 58--92.

\bibitem{lipatov}
Lipatov L.N., Divergence of the perturbation theory series and the
 quasiclassical theory, \textit{Sov. Phys. JETP} \textbf{45} (1977), 216--223.

\bibitem{Ma:1975vn}
Ma E., Asymptotic freedom and a `quark' model in six-dimension, \href{https://doi.org/10.1143/PTP.54.1828}{\textit{Progr.
 Theoret. Phys.}} \textbf{54} (1975), 1828--1832.

\bibitem{Macfarlane:1974vp}
Macfarlane A.J., Woo G., $\phi^3$ theory in six dimensions and the
 renormalization group, \href{https://doi.org/10.1016/0550-3213(74)90306-X}{\textit{Nuclear Phys.~B}} \textbf{77} (1974), 91--108,
 {E}rratum,
 \href{https://doi.org/10.1016/0550-3213(75)90361-2}{\textit{Nuclear Phys.~B}}
 \textbf{86} (1975), 548--548.

\bibitem{Mahmoud:2020vww}
Mahmoud A.A., Yeats K., Connected chord diagrams and the combinatorics of
 asymptotic expansions, \href{https://arxiv.org/abs/2010.06550}{arXiv:2010.06550}.

\bibitem{Maiezza:2018pkk}
Maiezza A., Vasquez J.C., Renormalons in a general quantum field theory,
 \href{https://doi.org/10.1016/j.aop.2018.04.027}{\textit{Ann. Physics}} \textbf{394} (2018), 84--97, \href{https://arxiv.org/abs/1802.06022}{arXiv:1802.06022}.

\bibitem{Marino:2012zq}
Mari\~no M., Lectures on non-perturbative effects in large {$N$} gauge
 theories, matrix models and strings, \href{https://doi.org/10.1002/prop.201400005}{\textit{Fortschr. Phys.}} \textbf{62}
 (2014), 455--540, \href{https://arxiv.org/abs/1206.6272}{arXiv:1206.6272}.

\bibitem{Marino:2019eym}
Mari\~no M., Reis T., Renormalons in integrable field theories, \href{https://doi.org/10.1007/jhep04(2020)160}{\textit{J.~High
 Energy Phys.}} \textbf{2020} (2020), no.~4, 160, 36~pages,
 \href{https://arxiv.org/abs/1909.12134}{arXiv:1909.12134}.

\bibitem{Marino:2019fvu}
Mari\~no M., Reis T., A new renormalon in two dimensions, \href{https://doi.org/10.1007/jhep07(2020)216}{\textit{J.~High
 Energy Phys.}} \textbf{2020} (2020), no.~7, 216, 34~pages,
 \href{https://arxiv.org/abs/1912.06228}{arXiv:1912.06228}.


\bibitem{Marino:2020dgc}
Mari\~no M., Reis T., Resurgence and renormalons in the one-dimensional Hubbard
 model, \href{https://arxiv.org/abs/2006.05131}{arXiv:2006.05131}.

\bibitem{Marie:2012cc}
Marie N., Yeats K., A chord diagram expansion coming from some
 {D}yson--{S}chwinger equations, \href{https://doi.org/10.4310/CNTP.2013.v7.n2.a2}{\textit{Commun. Number Theory Phys.}}
 \textbf{7} (2013), 251--291, \href{https://arxiv.org/abs/1210.5457}{arXiv:1210.5457}.

\bibitem{mckane-thesis}
McKane A.J., Vacuum instability in scalar field theories, Ph.D.~Thesis,
 {U}niversity of Southampton, 1978.

\bibitem{Mckane:1978me}
McKane A.J., Vacuum instability in scalar field theories, \href{https://doi.org/10.1016/0550-3213(79)90086-5}{\textit{Nuclear
 Phys.~B}} \textbf{152} (1979), 166--188.

\bibitem{McKane:2018ocs}
McKane A.J., Perturbation expansions at large order: results for scalar field
 theories revisite, \href{https://doi.org/10.1088/1751-8121/aaf768}{\textit{J.~Phys.~A: Math. Theor.}} \textbf{52} (2019),
 055401, 23~pages, \href{https://arxiv.org/abs/1807.00656}{arXiv:1807.00656}.

\bibitem{Misumi:2015dua}
Misumi T., Nitta M., Sakai N., Resurgence in sine-{G}ordon quantum mechanics:
 exact agreement between multi-instantons and uniform {WKB}, \href{https://doi.org/10.1007/JHEP09(2015)157}{\textit{J.~High
 Energy Phys.}} \textbf{2015} (2015), no.~9, 157, 41~pages,
 \href{https://arxiv.org/abs/1507.00408}{arXiv:1507.00408}.

\bibitem{mitschi2016divergent}
Mitschi C., Sauzin D., Divergent series, summability and resurgence.
 {I}.~Monodromy and resurgence, \textit{Lecture Notes in Math.}, Vol.~2153,
 \href{https://doi.org/10.1007/978-3-319-28736-2}{Springer}, Cham, 2016.

\bibitem{Panzer:2014caa}
Panzer E., Algorithms for the symbolic integration of hyperlogarithms with
 applications to Feynman integrals, \href{https://doi.org/10.1016/j.cpc.2014.10.019}{\textit{Comput. Phys. Commun.}}
 \textbf{188} (2015), 148--166, \href{https://arxiv.org/abs/1403.3385}{arXiv:1403.3385}.

\bibitem{sauzin}
Sauzin D., Introduction to 1-summability and resurgence, \href{https://arxiv.org/abs/1405.0356}{arXiv:1405.0356}.

\bibitem{Schnetz:2013hqa}
Schnetz O., Graphical functions and single-valued multiple polylogarithms,
 \href{https://doi.org/10.4310/CNTP.2014.v8.n4.a1}{\textit{Commun. Number Theory Phys.}} \textbf{8} (2014), 589--675,
 \href{https://arxiv.org/abs/1302.6445}{arXiv:1302.6445}.

\bibitem{Schnetz:2017bko}
Schnetz O., The {G}alois coaction on the electron anomalous magnetic moment,
 \href{https://doi.org/10.4310/cntp.2018.v12.n2.a4}{\textit{Commun. Number Theory Phys.}} \textbf{12} (2018), 335--354,
 \href{https://arxiv.org/abs/1711.05118}{arXiv:1711.05118}.

\bibitem{Schnetz:2016fhy}
Schnetz O., Numbers and functions in quantum field theory, \href{https://doi.org/10.1103/physrevd.97.085018}{\textit{Phys.
 Rev.~D}} \textbf{97} (2018), 085018, 20~pages, \href{https://arxiv.org/abs/1606.08598}{arXiv:1606.08598}.

\bibitem{oeis}
Sloane N.J.A., The on-line encyclopedia of integer sequences,
 \url{http://oeis.org/}.

\bibitem{stahl}
Stahl H., The convergence of {P}ad\'e approximants to functions with branch
 points, \href{https://doi.org/10.1006/jath.1997.3141}{\textit{J.~Approx. Theory}} \textbf{91} (1997), 139--204.

\bibitem{vanBaalen:2008tc}
van Baalen G., Kreimer D., Uminsky D., Yeats K., The {QED} {$\beta$}-function
 from global solutions to {D}yson--{S}chwinger equations, \href{https://doi.org/10.1016/j.aop.2008.05.007}{\textit{Ann.
 Physics}} \textbf{324} (2009), 205--219, \href{https://arxiv.org/abs/0805.0826}{arXiv:0805.0826}.

\bibitem{vanBaalen:2009hu}
van Baalen G., Kreimer D., Uminsky D., Yeats K., The {QCD} {$\beta$}-function
 from global solutions to {D}yson--{S}chwinger equations, \href{https://doi.org/10.1016/j.aop.2009.10.011}{\textit{Ann. Physics}}
 \textbf{325} (2010), 300--324, \href{https://arxiv.org/abs/0906.1754}{arXiv:0906.1754}.

\bibitem{Yeats:2008zy}
Yeats K., Growth estimates for {D}yson--{S}chwinger equations, Ph.D.~Thesis,
 {U}niversity of {W}aterloo, 2008, \href{https://arxiv.org/abs/0810.2249}{arXiv:0810.2249}.

\bibitem{yeats2017combinatorial}
Yeats K., A combinatorial perspective on quantum field theory,
 \textit{SpringerBriefs in Mathematical Physics}, Vol.~15, \href{https://doi.org/10.1007/978-3-319-47551-6}{Springer}, Cham,
 2017.

\bibitem{ZinnJustin:1979db}
Zinn-Justin J., Expansion around instantons in quantum mechanics,
 \href{https://doi.org/10.1063/1.524919}{\textit{J.~Math. Phys.}} \textbf{22} (1981), 511--520.

\bibitem{ZinnJustin:2002ru}
Zinn-Justin J., Quantum field theory and critical phenomena,
 \textit{International Series of Monographs on Physics}, Vol.~77, The
 Clarendon Press, Oxford University Press, New York, 1989.

\bibitem{ZinnJustin:2004ib}
Zinn-Justin J., Jentschura U.D., Multi-instantons and exact results.
 {I}.~{C}onjectures, {WKB} expansions, and instanton interactions,
 \href{https://doi.org/10.1016/j.aop.2004.04.004}{\textit{Ann. Physics}} \textbf{313} (2004), 197--267,
 \href{https://arxiv.org/abs/quant-ph/0501136}{arXiv:quant-ph/0501136}.

\bibitem{ZinnJustin:2004ib+}
Zinn-Justin J., Jentschura U.D., Multi-instantons and exact results. {II}.~{S}pecific cases, higher-order effects, and numerical calculations,
 \href{https://doi.org/10.1016/j.aop.2004.04.003}{\textit{Ann. Physics}} \textbf{313} (2004), 269--325.

\end{thebibliography}
\end{document}